\documentstyle[eqsecnum,multicol,epsfig,aps,prl,array]{revtex}
\newcommand{\bleq}{\ifpreprintsty
                   \else
                   \end{multicols}\widetext \vspace*{-3.5ex}{\tiny
                   
		\noindent\begin{tabular}[t]{c|}
                   \parbox{0.493\hsize}{~} \\ \hline \end{tabular}}
	                              \fi}
\newcommand{\eleq}{\ifpreprintsty
                   \else
                   {\tiny\hspace*{\fill}\begin{tabular}[t]{|c}\hline
                    \parbox{0.49\hsize}{~} \\
                    \end{tabular}}\vspace*{-2.5ex}\begin{multicols}{2}
	            \narrowtext
                    \fi}
\newcommand{\bcols}{\ifpreprintsty\else\begin{multicols}{2} 
	\narrowtext\fi}
\newcommand{\ecols}{\ifpreprintsty\else\end{multicols}\fi}

\draft
\widetext
\begin{document}
\title{Droplets nucleation, and Smoluchowski's equation 
with growth and injection of particles} 
\author{St\'ephane Cueille and Cl\'ement Sire}
\address{Laboratoire de Physique Quantique (UMR C5626 du CNRS),
Universit\'e Paul Sabatier\\
31062 Toulouse Cedex, France.\\}

\date{\today}
\maketitle
\begin{abstract}
We show that  models for homogeneous and heterogeneous nucleation of
$D$-dimensional droplets in a $d$-dimensional medium  
are described in mean-field by a modified Smoluchowski equation  for the
distribution $N(s,t)$ of droplets   masses $s$, with
additional terms accounting for exogenous growth from vapor absorption,
and injection of small droplets when the model allows renucleation. The
corresponding collision kernel is derived in both cases.
For a generic collision kernel $K$, the equation describes a clustering
process with  clusters of mass $s$ growing between collision with
$\dot{s}\propto s^\beta$, and injection of monomers   at a rate $I$. 
General properties of this equation are studied. The gel criterion is 
determined. Without injection, exact solutions are found with a 
constant kernel, exhibiting unusual scaling behavior. For a general 
kernel, under the scaling assumption $N(s,t)\sim Y(t)^{-1}f(s/S(t))$, we 
determine the asymptotics of $S(t)$ and $Y(t)$, and derive the scaling 
equation. Depending on $\beta$ and  $K$,
 a great diversity of behaviors is
found.  For constant injection, there is an asymptotic steady state with
 $N(s,t=\infty)\propto s^{-\tau}$ and $\tau$ is determined. 
The case of a constant mass injection rate is related to homogeneous nucleation and is
studied. Finally, we show how these results shed some new light on 
heterogeneous nucleation with $d=D$. For $d=D=2$ (discs on a plane), numerical
simulations are performed, in good agreement with the mean-field results.   
\end{abstract}
\bcols 
\section{Introduction}
Some very simple and practically important physical phenomena can be related
to aggregation models, and as a consequence a host of experimental, numerical
and theoretical studies are to be found in the literature. Classical fields
of application are  atmosphere sciences, material sciences, chemical engineering  and cosmology, 
among others \cite{friedlander77,family84,stanley86,vicsek92}.
 One of these 
phenomena is 
{\it dropwise condensation} on a substrate \cite{meakindrop92},
 for instance water on a cold window pain,  which bears on 
 important implications in heat transfer engineering and material sciences,
and generates fascinating droplets patterns, also called {\it breath figures}
\cite{beysens86}.    
Droplets grow from vapor, and when two droplets 
come into contact, they coalesce to form a bigger droplet, 
with mass (or volume) conservation. The underlying physics is rich, and it is not our
purpose to fully discuss it here (see \cite{meakindrop92}). Simple 
models have been introduced to describe the late stage of droplet growth
 and coalescence, of basically two kinds
\cite{meakindrop92,family88,family89,fritter91,derrida91,marcos95}. 
On the one hand, in heterogeneous nucleation models, one starts from a fixed number of nucleation sites
 (in physical situations
these might be dust particles, substrate defects...). Droplets grow on these
sites through vapor absorption, and when two neighboring droplets overlap,
they coalesce to form a single droplet, thus reducing the number of droplets.
On the other hand, in homogeneous growth models \cite{family88}, nucleation can occur anywhere
on the substrate: some very small droplets are randomly deposited, 
which leads to the growth of existing droplets, and the creation of new 
small droplets if  deposition occurs in a free zone.

Both kinds of models are aggregation models, with the unusual feature 
that  the aggregating particles
grow between coalescence events. Homogeneous growth has the additional
feature that there is a source of small droplets, which relates this
model to aggregation  in presence of a source, which is also  widely-studied
 in the literature 
\cite{klett75,white82,crump82,hendriks85,vicsek85,racz85,hayakawa87,takayasu88,takayasu89,majumdar93}.
 Experiments and numerical simulations
 \cite{meakindrop92,family89} show that 
the time dependent droplet mass distribution $N(s,t)$ ($s$ being the volume 
of the droplets), exhibits dynamic scaling, i.e. that for  large
times $N(s,t)\propto S(t)^{-\theta}f(s/S(t))$. $S(t)$ is a typical droplet
mass (proportional to  $<s^2>/<s>^2$) and has a power law divergence at large 
time $S(t)\propto t^z$. $\theta$ and $z$ are  dynamic exponents that do not
depend on the fine details of the model but only on its main features, 
such as its conservation laws. In heterogeneous growth models, 
the scaling function $f(x)$ is narrow-shaped,
whereas in homogeneous growth one observes the superposition 
of a monodispersed 
distribution of large droplets (with masses of the same order as $S(t)$),
and a polydispersed distribution of small droplets with a power law divergence
of the scaling function $f(x)\propto x^{-\tau}$ at small $x$.
 $\tau$ is nontrivial and
less than $\theta$ \cite{family89}. Such nontrivial polydispersity
exponents frequently appear  in aggregation models. For usual models (without 
injection or growth), they can be accounted for by Smoluchowski's
 mean-field approach  \cite{smoluchowski18}: neglecting fluctuations and 
multiple collisions, one
can write down a rate equation,
\begin{eqnarray} \label{smoleq}
\partial_t N(s,t)
&=\frac{1}{2}\int N(s_1,t)N(s-s_1,t)K(s_1,s-s_1)\,ds_1\nonumber\\
&- N(s,t) \int N(s_1,t) K(s,s_1) \,ds_1,
\end{eqnarray} 
where the collision kernel $K(s_1,s_2)$ is the probability of a coalescence
event between two droplets of masses $s_1$ and $s_2$. Smoluchowski's approach
is valid above an upper critical dimension which is often $2$, but is in
principle model-dependent \cite{vdg89}. Van Dongen and Ernst \cite{vdg85prl}, 
classified the kernels according to their homogeneity and asymptotic 
behavior:
\begin{eqnarray}\label{genscal1}
K(bx,by)=b^\lambda K(x,y),\\
K(x,y)\sim x^\mu y^\nu \,\,\, (y \gg x). \label{genscal2}
\end{eqnarray} 
Nontrivial polydispersity exponents appear in the case $\mu=0$ 
\cite{vdg85prl,vdg87scal,nontriv97}, whereas
for $\mu>0$, $\tau$ is equal to $1+\lambda$, and for $\mu<0$ the distribution
is bell-shaped. 

In this article we are concerned with studying the mean-field Smoluchowski
approach to those aggregation processes, such as droplet nucleation, where
 individual particles or clusters grow between collisions (with a growth law 
$\dot{s}\propto s^\beta$), and/or  monomers are injected with a possibly 
time-dependent injection rate. Our first motivation to this study is a better
understanding of droplets
deposition, growth and coalescence models, and throughout the article, 
 the example of both
homogeneous and heterogeneous droplet nucleation will be used as 
an illustration of the results. Certain results presented here
have  already appeared in a summarized form in \cite{droplett}.    

Sec. \ref{derive} discusses  nucleation models introduced by Family and Meakin 
\cite{family89}, and derives the corresponding Smoluchowski equation under
the mean-field assumption.   This  leads us to a generalized Smoluchowski equation with an additional
 exogenous growth
term  $\partial_s (s^\beta N)$ in the left hand side of Eq. 
(\ref{smoleq}) and a time dependent source term $I(t)\delta(s-s_0)$ in its 
right hand side.

Sec. \ref{sec:smol} is a general study of the extended Smoluchowski equation
corresponding to aggregation with exogenous growth and injection,
  such as the one
obtained in Sec. \ref{derive}, with a generic homogeneous kernel. The
gelation criterion is investigated, depending on  $\lambda$ defined in
 Eq. (\ref{genscal1}). It is found that the system is nongelling for
$\mbox{max}(\lambda,\beta)\leq 1$. Some exact solutions are found in the absence of
injection ($K=1$, with $\beta=0$ or $\beta=1$). The scaling properties of
the equation without injection are investigated. It is found that the scaling 
behavior depends on $\beta$, $\lambda$ and $\mu$, and is richer than for 
standard Smoluchowski's equation. The occurrence of polydispersity exponents
is discussed. When polydispersity occurs, the scaling equation is the same
as for standard Smoluchowski's equation, and the methods recently introduced
by the present authors \cite{nontriv97} can be used to compute nontrivial 
$\tau$ exponents. 

For constant injection of monomer, it is found that the distribution reaches
 at infinite time a polydispersed steady state with a power law 
large $s$ decay, with $\tau=(3+\lambda)/2$ if $\beta<(1+\lambda)/2$,
and $\tau=2+\lambda-\beta$, if $\beta>(1+\lambda)/2$. Then we consider 
a constant mass injection rate, with a self-consistent $I(t)$, 
for $\lambda=2\beta-1$,
which is relevant to homogeneous nucleation,
and we show that the injection rate $I(t)$ is vanishing, in agreement with 
the droplets model. We also investigate scaling solutions, and suggest that
including pair correlations may be necessary to find a consistent scaling for 
homogeneous growth.

Sec. \ref{heteropoly} is an application of the scaling mean-field  results
 to heterogeneous
growth with $d=D$. Droplets radii $r=s^{1/D}$ grow as $\dot{r}\propto
 r^\omega$ ($\beta=1+(\omega-1)/D$), and 
polydispersity with a nontrivial $\tau$ 
occurs for $\omega\geq 0$, while the scaling function is monodipersed for
$\omega<0$. 
Mean-field polydispersity exponents are
computed using the variational method introduced in \cite{nontriv97}, and
described in Sec. \ref{sec:smol}.
 Numerical results for the scaling function  are in
qualitative agreement with mean-field results, and the expected cross-over
from monodispersity to polydispersity at $\omega=0$ is observed.

\section{Droplets deposition, growth and coalescence in mean-field}
\label{derive}
As mentioned in the introduction, interest in droplet nucleation computer
models was primarily aroused by practical applications in heat transfer
 engineering (see references in \cite{meakindrop92}). In the last ten year,
 however, and since the seminal  work of Beysens and Knobler
\cite{beysens86}, the focus was set on the formation of breath figures 
(see Fig. \ref{fig:iconf}), with computer models aimed to 
study the kinetics of the droplet mass distribution 
\cite{family88,family89,fritter91,meakindrop92}, the asymptotic surface (or
line) coverage \cite{derrida91}, or the time evolution of the ``dry'' fraction
(the surface fraction which has never been covered by any droplet)
\cite{marcos95}. 

In this article, we shall consider 
the specific models introduced by Family and Meakin
\cite{family88,family89}, 
both for homogeneous and heterogeneous nucleation. We shall now describe
these models and derive the corresponding Smoluchowski equation. These
equations are special cases of a generalized Smoluchowski equation which
will be studied in Sec. \ref{sec:smol}.

\subsection{Homogeneous nucleation}
First, let us describe homogeneous nucleation, or the {\it deposition and
 coalescence}
model. Between $t$ and $t+\delta t$
a small droplet of mass $s_0$ is randomly deposited on the $d$-dimensional substrate where
it forms a spherical cap with radius $s_0^{1/D}$. 
If it overlaps an existing droplet of mass $s$, they coalesce to form a new
 droplet with mass $s+s_0$ and radius $(s+s_0)^{1/D}$, centered at the 
center of mass of the two coalescing droplets. If the new droplet overlaps a
surrounding droplet, they coalesce with the same rule, and so on.
The distribution of droplet  masses  $N(s,t)$ exhibits dynamic 
scaling,
\begin{equation}\label{dynscal}
N(s,t)\sim S(t)^{-\theta}f(s/S(t)).
\end{equation}
The
typical mass scale $S(t)$ can be defined by,
\begin{equation}
S(t)=\frac{<s^2>}{<s>}=\frac{\int s^2N(s,t)}{\int sN(s,t)}.
\end{equation}
 The dynamical exponents $\theta$ and $z$ are easily determined from 
physical arguments \cite{meakindrop92,family89}. 
 
Since the mass injection rate is constant and the mass is conserved in the
coalescence process, we must have,
\begin{equation}
t \propto \int_0^{+\infty}\!\!sN(s,t)dt\propto S(t)^{2-\theta} 
\int_0^{+\infty}\!\!
xf(x)dx,  
\end{equation} 
which, from the definition of $z$, implies the scaling law $z(2-\theta)=1$.
Then we note that the fraction of substrate ``area'' occupied by
the droplets is,
\begin{equation}
\int_0^{+\infty}s^\frac{d}{D}N(s,t)ds \propto S(t)^{1+\frac{d}{D}-\theta},
\end{equation}  
and cannot diverge or vanish, so that $\theta=1+d/D$. From the scaling
law, we get $z=D/(D-d)$. 

The scaling behavior of the total number of droplets $n(t)$ depends on the 
small $x$ behavior of the scaling function $f(x)$.  If the scaling function 
is integrable in zero, it is easily seen that $n(t)\propto S(t)^{1-\theta}$,
whereas if $f(x)\propto x^{-\tau}$ with $\tau>1$, 
$n(t)\propto S(t)^{\tau-\theta}$. 

\begin{figure}
\begin{center}
\epsfig{figure=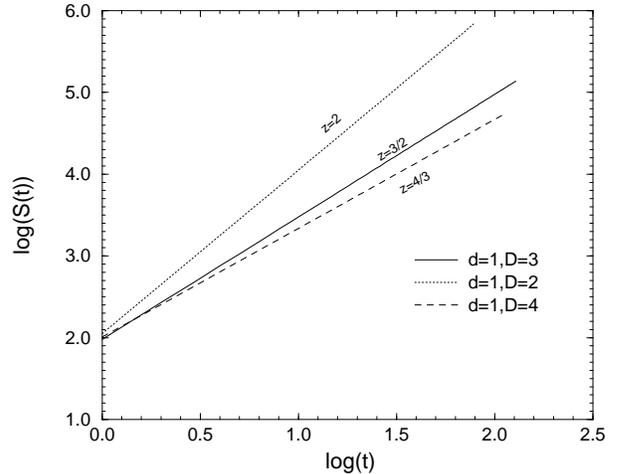,width=\linewidth}
\caption{$S(t)$ from numerical simulations of deposition and coalescence
in $d=1$. During one time step, $10^7$ droplets of radius $0.75$ were
deposited on a line of length $5\times 10^5$. Ten samples were averaged to
obtain these data. The scaling laws $S(t)^z$ with $z=D/(D-1)$ are very well
obeyed.}
\label{fig:growthlaw}
\end{center}
\end{figure}

Numerical results as obtained in \cite{family89} confirm the scaling
hypothesis, and the theoretical values of $\theta$ and $z$. Actually, authors
found some values of $z$ a few percents smaller than $D/(D-d)$, since
 simulations did not reach sufficiently large times. To illustrate 
our discussion, we present here our own numerical
simulations in one dimension, for $D=2,3,4$.
 Droplets of radius $r_0=0.75$ were randomly deposited on 
a one dimensional ``substrate'' of length $L=5\times 10^5$. 
During one simulation 
time step, about $10^7$ droplets were deposited. 
Ten samples were averaged to obtain
the final data. Fig. \ref{fig:growthlaw} shows the evolution of $S(t)$.
We observe  
a  power law $S(t)\propto t^z$ with numerical values of $z$ equal to
their theoretical values with excellent accuracy 
(for instance for $D=3$ we find $z=1.4995\pm
0.00015$ to be compared with the theoretical $z=3/2$). 

\begin{figure}
\begin{center}
\epsfig{figure=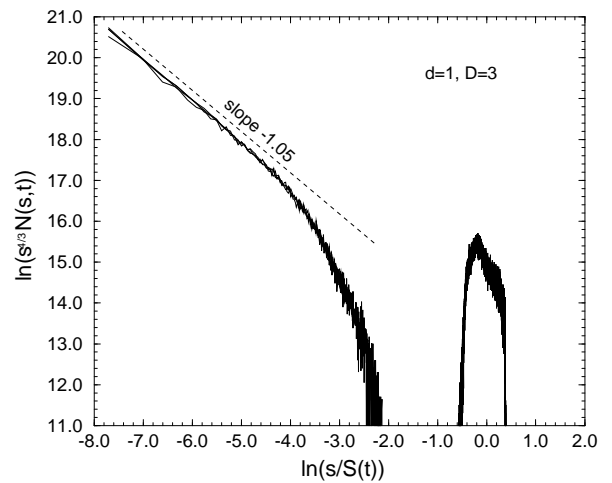,width=\linewidth}
\caption{Scaling of the mass distribution $N(s,t)$ for droplet deposition
with $d=1$ and $D=3$. The picture shows the excellent data collapse, with the
theoretical value $4/3$ for $\theta$, of the distribution at four different
times when $S$ has respectively reached the values $6053$, $11116$, $15539$,
and $17112$. The scaling function is composed of a polydispersed
contribution
of small droplets and a monodispersed contribution of droplets of mass of
order $S(t)$. }
\label{fig:distri}
\end{center}
\end{figure}

\begin{figure}
\begin{center}
\epsfig{figure=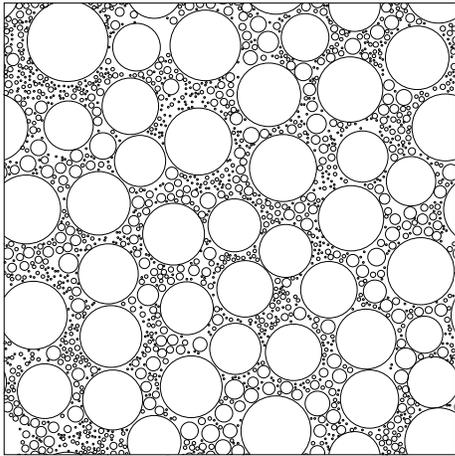,width=0.7\linewidth}
\caption{A typical configuration in the scaling regime of deposition and
coalescence of three-dimensional droplets on a plane. The picture shows a
configuration after that $655360$ droplets of radius $0.75$ were deposited
on a $256\times 256$ square surface with periodic boundary conditions, and 
$S(t)$ has reached the value $10348.9$. Two distinct populations of droplets
are clearly visible: a monodispersed population of large droplets and a
population of small droplets with a broad distribution.}
\vspace{0.2cm}
\label{fig:iconf}
\end{center}
\end{figure}

The scaling for the distribution function $N(s,t)$ is illustrated 
in Fig. \ref{fig:distri} for $D=3$. The data collapse is obtained
with the theoretical value $\theta=4/3$.
The scaling function  is composed of two distinct parts: a polydispersed
small droplets distribution, with a small argument divergence of the scaling
 function associated with an exponent $\tau$ bigger than $1$, 
well separated from a bell-shaped monodispersed
distribution of bigger droplets centered around $s=S(t)$. Most of the
droplets in the system contribute to the small droplets distribution,
which determines, since $\tau>1$, the behavior of $n(t)$, whereas the
population of bigger droplets contains most of the mass, and
$S(t)$ is the typical mass of big droplets. The two distinct populations of
 droplets are clearly
visible on Fig. \ref{fig:iconf}, which shows a typical configuration of
droplets obtained by simulation of the deposition and coalescence model 
for $d=2$ and $D=3$ (spherical droplets on a plane). As shown in 
\cite{family89} the obtained droplets patterns are qualitatively very close 
to the one obtained in some experiments of vapor deposition of thin films.

\begin{figure}
\begin{center}
\epsfig{figure=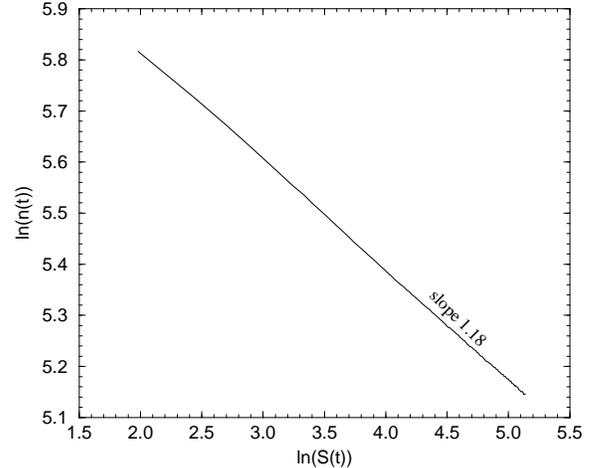,width=\linewidth}
\caption{The scaling law $n(t)\propto S(t)^{\tau-\theta}$ for $\tau>1$
is very well obeyed in numerical simulations in $d=1$. The figure is a
plot of $\log(n)$ versus $\log(S)$ for $D=3$, for which we find
$\tau=1.18\pm 0.03$.}
\label{fig:tau}
\end{center}
\end{figure}

The exponent $\tau$ can be determined from the numerical determination
of the scaling function, but with important uncertainty due to
statistical limitations. Thus, a better method is to extract $\tau$ from 
$n(t)\propto (S(t))^{(\tau-\theta)}$. This power law behavior is
well recovered in numerical simulation (see Fig. \ref{fig:tau}), and
gives the results, $\tau=1.264\pm 0.002$ ($D=2$),
 $\tau=1.18\pm 0.03$ ($D=3$), $\tau=1.074\pm 0.001$ ($D=3$).

 Direct determination
from the scaling function yields $\tau \approx 1.05$ for $D=3$, 
which is only in fair
agreement with the previous result, but  this underestimated value of $\tau$
would be improved by using larger, or more numerous samples.
The exponent $\tau$ is strictly less than $\theta$, but is nontrivial
and cannot be determined by simple physical arguments.

Therefore, it is interesting to derive a Smoluchowski equation for this model,
and check the mean field value of $\tau$, if possible. 
Family and Meakin \cite{family89} showed from scaling arguments that the
 coalescence kernel should have a homogeneity $\lambda=2d/D-1$, but they 
did not determine its specific form.  We proceed now to the determination 
of the equation. Neglecting multiple collisions, we examine the different 
events affecting the distribution $N(s,t)$.

\bleq

Between $t$ and $t+1$,  a droplet of radius $s=ks_0$ is created as an
outcome of the following processes:
\begin{enumerate}
\item a droplet of mass $s_0$ falls on a droplet of mass $s_1\leq s-s_0$, 
 which occurs with probability  $\Omega_1(s_1^{1/D}+s_0^{1/D})^d$,
$\Omega_1$ being a mass independent geometric factor.
 The droplet of mass $s_1$ 
consequently reaches a mass $s_1+s_0$ 
 Then it coalesces with a neighboring
droplet of mass $s_2=s-s_1-s_0$ provided that they interpenetrate. 

The number of such events is,
\begin{equation}
  \Omega_1\Omega_2N(s_1)N(s_2)(s_1^{\frac{1}{D}}+s_0^\frac{1}{D})^d 
 \int_0^{(s_1+s_0)^\frac{1}{D}-s_1^\frac{1}{D}}
G(s_1,s_2,r,t) (s_1^\frac{1}{D}+s_2^\frac{1}{D}+r)^{d-1} \,dr
\end{equation}
($\Omega_2$ is another geometric factor).
$G(s_1,s_2,r,t)$ is the probability density that
a given droplet of mass $s_1$ has a droplet of mass $s_2$ at distance 
$s_1^\frac{1}{D}+s_2^\frac{1}{D}+r$ as first neighbor.

\item 
a droplet of mass $s_0$ falls on a droplet of mass $s-s_0$ with which 
it coalesces, and the obtained droplet does not overlap any other droplet,

\begin{eqnarray}
 &\mbox{number of events} =\Omega_1 N(s-s_0)((s-s_0)^\frac{1}{D}+s_0^\frac{1}{D})^d\nonumber\\
&\times \left(1-\Omega_2\sum_{s_1=k_1s_0}N(s_1)\int_0^{s^\frac{1}{D}-(s-s_0)^\frac{1}{D}}
 G(s-s_0,s_1,r,t) ((s-s_0)^\frac{1}{D}+s_1^\frac{1}{D}+r)^{d-1} \,dr\right).
\end{eqnarray}

\item a droplet falls in an empty space between droplets,
  \begin{equation}
   \mbox{number of events}\propto(1-\phi(t)) \delta_{s,s_0},
  \end{equation}
where $(1-\phi(t))$ is the empty area fraction.
\end{enumerate}

A droplet of radius $s$ disappears due to the following events:

\begin{enumerate}
\item it coalesces with a droplet of radius $s_1+s_0$ which has grown

\begin{equation}
\mbox{number of events}=\Omega_1\Omega_2 N(s)N(s_1)(s_1^\frac{1}{D}+s_0^\frac{1}{D})^d \int_0^{(s_1+s_0)^\frac{1}{D}-s_1^\frac{1}{D}}
G(s_1,s,r,t) (s_1^\frac{1}{D}+s^\frac{1}{D}+r)^{d-1} \,dr.
\end{equation}

\item it grows:
\begin{equation}
\mbox{number of events}\propto N(s)(s^\frac{1}{D}+s_0^\frac{1}{D})^d.
\end{equation}
\end{enumerate}

To describe the large time scaling regime, we can  take the continuous 
limit of small (but finite) $s_0$, to obtain
the following  continuous kinetic equation,
\begin{eqnarray}\label{homosmolu}
\partial_t N(s,t)+\partial_s (s^\frac{d}{D}
N(s,t))&=& \frac{1}{2} \int_{0}^{s}  N(s_1,t)N(s-s_1,t)K(s_1,s-s_1,t) ds_1\\
& & -N(s,t)\int_{0}^{+\infty}N(s_1,t)K(s,s_1,t) ds_1 +I(t) \delta(s-s_0),
\end{eqnarray}
where the symmetric kernel $K(x,y,t)$ is,
\begin{equation}
K(x,y,t)=\lim_{\varepsilon \to 0} \frac{x^\frac{d}{D}}{\varepsilon}\int_0^{(x+\varepsilon)^\frac{1}{D}-x^\frac{1}{D}}
G(x,y,r,t) (x^\frac{1}{D}+y^\frac{1}{D}+r)^{d-1} \,dr + \mbox{\rm symmetric}.
\end{equation}
The time and  mass units were
redefined to eliminate   multiplicative constants in the equation.
$I(t)$ is consequently renormalized  to $I(t)=c (1-\phi(t))$,
 where $c$ is a constant which could be easily determined, but is not 
essential 
to our discussion. It should be noticed that the distribution function is
zero below $s_0$ at any $t$.
\eleq

The mean-field approximation consists in neglecting spatial correlations
{\it i.e.}
in  taking $G(x,y,r,t)=1$. We get:
 \begin{equation}\label{depkern}
   K(x,y,t)=(x^{\frac{d+1}{D} -1}+y^{\frac{d+1}{D}-1})(x^\frac{1}{D}+y^\frac{1}{D})^{d-1}.  
 \end{equation}

This kernel has the homogeneity $\lambda=2d/D-1$ derived from scaling 
arguments by Family and Meakin \cite{family89}. Eq. (\ref{homosmolu})
is not a standard Smoluchowski equation since it incorporates two additional
terms: an {\it exogenous  growth} term $\partial_s (s^\frac{d}{D}
N(s,t))$, describing intercollision growth of droplets through absorption
of small droplets, and a time dependent {\it injection} term.  Moreover, the
injection term is  {\it self-consistent}:
 being proportional to the free surface 
 fraction, it is
a functional of $N(s,t)$ (see Sec. \ref{constmass}).

\begin{figure}
\begin{center}
\epsfig{figure=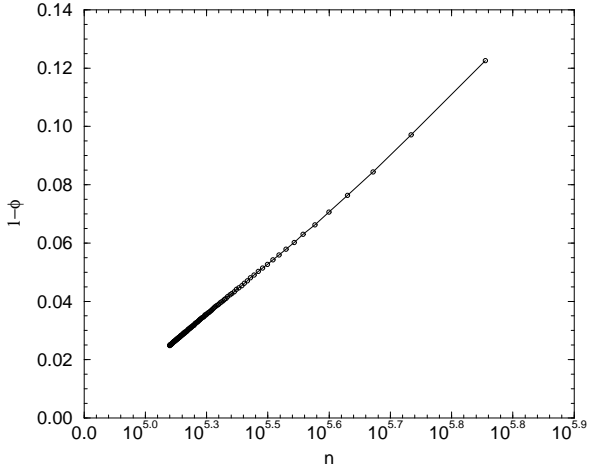,width=\linewidth}
\caption{Plot of the free substrate area $1-\phi(t)$ versus the number of
droplets $n(t)$ for a simulation of homogeneous growth with $d=1$ and $D=3$. At
large time (small $n$), the two quantities are proportional,
 as understood in the framework of the scaling theory.}
\label{fig:area}
\end{center}
\end{figure}

\vspace{0.5cm}
$I(t)$ is vanishing at large time, since the  
surface fraction covered by droplets goes to one in homogeneous growth processes,
through renucleation in empty spaces (in heterogeneous nucleation models, the
coverage goes to a value $\bar{\phi}<1$).  
In fact, our numerical simulations in 1D
show that in the scaling regime $1-\phi(t)\propto n(t)$, as
illustrated in Fig. \ref{fig:area}.
 This result is easily recovered
from the scaling theory, 
\begin{equation}
 1-\phi(t)=1- \Omega S(t)^{(1+\frac{d}{D}-\theta)}
\int_{\frac{s_0}{S(t)}}^{+\infty}
x^\frac{d}{D}f(x) dx,
\end{equation} 
where $\Omega$ is a geometric constant factor,
which implies, since $\phi(t)\to 1$ and $\theta=1+d/D$, that
$\int_0^{+\infty} x^{d/D} f(x)=\Omega^{-1}$. Since 
$f(x)\propto x^{-\tau}$, we see that, 
$(1-\phi(t)) \propto(s_0/S(t))^{1+d/D-\tau}$, if $\tau>1$,
and $(1-\phi(t))\propto (s_0/S(t))^{d/D}$, if $\tau <1$, which yields
$(1-\phi(t))\propto n(t)$.  
 
\subsection{Heterogeneous nucleation}\label{droplets}

\begin{figure}
\begin{center}
\epsfig{figure=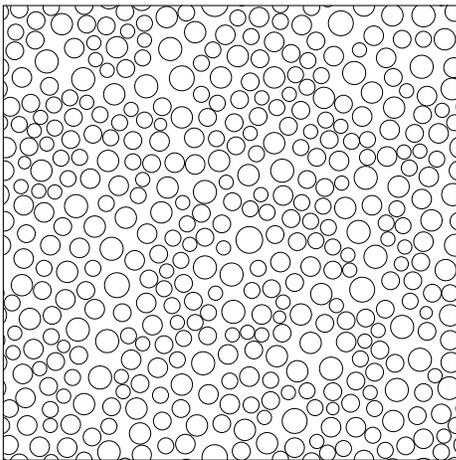,width=0.7\linewidth}
\caption{A typical configuration in the scaling regime of 
heterogeneous nucleation of droplets in $d=2$. The simulation was performed
for $\omega=-1$ on a $512\times 512$ square surface 
 with $256^2$ droplets of radius
$0.7$ in the initial condition.}
\label{fig:hetero3D}
\end{center}
\end{figure}
Heterogeneous nucleation \cite{meakindrop92} corresponds to the case,
common  for water vapor condensation,
when impurities on the substrate play a major role in 
droplet nucleation. A daily life example would be water condensation on a
dusty pain.   Nucleation occurs only on some
nucleation centers, existing droplets grow from vapor, and coalesce when
coming into contact,  but no new droplet 
can nucleate in empty spaces.  

In the {\it growth and coalescence} model introduced by Family and Meakin
\cite{family89},  one starts from an initial population of droplets of same
radius with no
 overlap. 
In the dynamics, individual droplets grow between collisions with,
\begin{equation}\label{drdt}
\dot{r}=A r^\omega,
\end{equation}
or, equivalently ($s=r^D$),
\begin{equation}
\dot{s}=D A s^\beta,
\end{equation}
with $\beta=(\omega+D-1)/D$. 
 In the following theoretical discussion
 we shall always set $A=1/D$, but in  numerical simulations $A$ was set
to $1$ (this just corresponds to a change in the time unit).
We must have $\omega \leq 1$ (or equivalently $\beta\leq 1$), otherwise
the system is gelling as the mass of an individual droplet growing without 
collision diverges at finite time.
One  step  of simulation consists in  increasing the radii of all
droplets  according to a discretization of 
Eq. (\ref{drdt}), then tracking down and resolving all the resulting 
coalescence events, with the same rules as for homogeneous nucleation.

 Fig. \ref{fig:hetero3D} is a snapshot   of a typical simulation of 
growth and coalescence of three dimensional droplets ($D=3$) on a plane
($d=2$). The simulations were carried out for $\omega=-1$ on a 
$512\times 512$ square surface with periodic boundary conditions from 
an initial population of $256^2$ droplets of radius $0.7$ randomly placed
without overlap. Qualitatively similar results are obtained when varying
$\omega$ \cite{family89}. The droplets configuration in
Fig. \ref{fig:hetero3D} appears  to be visually very different from the 
one in Fig. \ref{fig:iconf} for homogeneous nucleation. 
In heterogeneous growth,
there is  a single,  monodispersed population of droplets. Another
interesting feature of heterogeneous growth is that the surface coverage
tends to a limit $\bar{\phi}<1$ (\cite{vincent71,family89} and references in 
\cite{meakindrop92}), which depends on $D$ but not significantly on
$\omega$. The value of the asymptotic coverage was computed
by Derrida et al. \cite{derrida91} 
for several simplified models of coalescence in one dimension. Vincent 
\cite{vincent71}, derived $\bar{\phi}=0.57$ from an approximate  log-normal
scaling solution to a Smoluchowski mean-field equation (read below) 
with four-body collisions 
included  for $d=2$, $D=3$ and
$\omega=-2$, in excellent agreement with the numerical value 
$\bar{\phi}=0.55$.

\begin{figure}
\begin{center}
\epsfig{figure=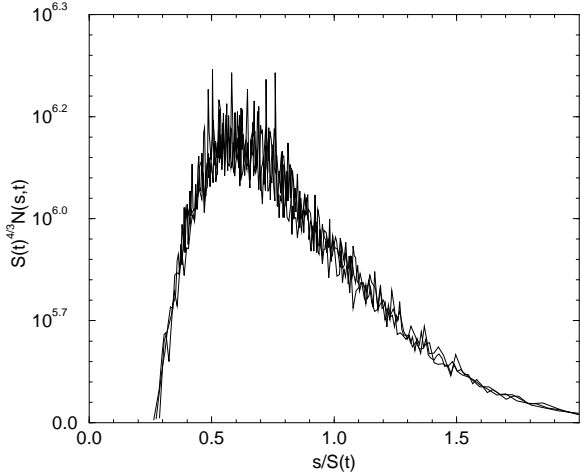,width=\linewidth}
\caption{A typical scaling function for heterogeneous nucleation with $d<D$.
The figure shows the data collapse of $N(s,t)$ in a simulation of growth and
coalescence with $\omega=-1$, $d=2$ and $D=3$. The scaling function is
clearly monodispersed and vanishes at $x_0\approx 0.2$.}
\label{fig:distri3D}
\end{center}
\end{figure}

Consistently  with Fig. \ref{fig:hetero3D}, the droplets mass distribution is  bell-shaped \cite{family89},
 as shown in Fig. 
\ref{fig:distri3D} for  $\omega=-1$. 
As in deposition and coalescence, an asymptotic scaling regime is reached 
at large time, as assessed by the data collapse obtained 
 for the mass distribution $N(s,t)$ at three
different times. The scaling form of Eq. (\ref{dynscal}) was used with 
 the theoretical value $\theta=1+d/D$, derived, as in the
 homogeneous case, from the fact that the surface coverage tends to a
constant.

The exponents $z$ can be determined from the fact that
$S(t)$ is the only mass scale in the asymptotic regime, which 
implies that the distance between 
droplets $n^{1/d}$ scales as $S^{1/D}$. Then, from a rough
 evaluation of the total collision rate, it is justified in
\cite{meakindrop92}
 that the growth law of the typical  droplet mass in the asymptotic regime
 is the same
as that of an individual droplet in the absence of collision,
except for a multiplicative constant renormalizing the growth rate, 
\begin{equation}
\dot{S}\propto  S^{\beta},
\end{equation}  
which leads to,
\begin{equation}
z=\frac{1}{1-\beta}=\frac{D}{1-\omega}.
\end{equation}
 These scaling results will be established 
for the corresponding Smoluchowski equation in Sec. \ref{sec:smol}.

A corollary of this scaling behavior is that the scaling function $f(x)$
 cannot be diverging at small $x$. 
A simple argument even shows that $f(x)$ is strictly zero below  a
finite $x_0>0$, as can be seen on Fig. 
\ref{fig:distri3D} (see also Fig. 2 in \cite{derrida91}).
In homogeneous nucleation, the lower cut-off of the distribution is the 
mass $s_0$ of the deposited droplets, which is time-independent, while
$S(t)$ diverges. Consequently, the lower cut-off of the scaling function is $0=\lim
s_0(t)/S(t)$, and $f(x)$ can have a power law divergence when $x\to 0$.
The situation is very different in heterogeneous nucleation.
 Consider the smallest droplets surviving at $t$. These are 
descendants of the droplets in the initial condition,
that have not experienced any collision since $t=0$. As a consequence,
the mass of the smallest surviving droplets $s_0(t)$ is, for a strictly 
monodispersed initial condition $N(s,0)\propto \delta(s-s_0)$,
\begin{equation}\label{s0}
s_0(t)\propto (s_0(0)^{1-\beta}+(1-\beta)t)^\frac{1}{1-\beta},
\end{equation}
and $s_0(t)/S(t)$ approaches a constant value $x_0>0$, independent on $s_0$,
when $t\to \infty$. Since $N(s,t)=0$ for $s<s_0(t)$, we see that 
$f(x)=0$ for $x<x_0$. 
 Figure \ref{fig:distri3D} shows the scaling function obtained from 
simulations for $d=2$, $D=3$ and $\omega=-1$. Numerical results give 
$S(t)\sim 16\, t^{1/(1-\beta)}$, with $\beta=1/3$, whereas $s_0(t)\sim
((1-\beta)Dt)^{1/(1-\beta)}$ from Eq. (\ref{s0}) ($D$ appears in the formula
since $A=1$ in the simulation), which leads to $x_0\approx 0.2$. This 
value of $x_0$ is fully consistent with  Fig. \ref{fig:distri3D}.

 However, if $\beta=1$, because the growth
of $s_0(t)$ is exponential in time, the situation can be  different. The 
collisions renormalize the growth of $S(t)$ and we expect at large time,
\begin{equation}\label{rengro}
\dot{S}\sim (1+\epsilon(t))S(t),
\end{equation}
where $\epsilon$ is strictly positive, and 
might tend to zero, but there is no
reason to expect that it should vanish so fast that $S(t)\propto s_0(t)$, 
and $S(t)$ can be much bigger than $s_0(t)$, leading to $x_0=0$, and a
possibly polydispersed distribution. This behavior will be found in Sec. \ref{sec:smol} for 
the corresponding Smoluchowski  equation.   
 We shall see that $S(t)$ is also much bigger
than $s_0(t)$ {\it in mean-field} in the case $d=D$,
 as $S(t)\propto (t\ln t)^{1/(1-\beta)}$,
and that polydispersity can occur in this case. 
Actually, 
Family and Meakin \cite{family89} observed 
a qualitatively different, broader mass
distribution, with polydispersity, in numerical simulations for $d=D$.
This case will be further investigated in Sec. \ref{heteropoly}.

Now we proceed  in deriving Smoluchowski's equation for this problem.
In an early numerical
and theoretical work, Vincent \cite{vincent71}, considered  heterogeneous
growth of three-dimensional droplets on a two-dimensional substrate.
 His numerical simulations concerned
early stages of growth (because he could not reach the large time asymptotic
regime), when the number of droplets decays exponentially in time, in contrast
with the power law behavior in the scaling regime, but he wrote down a 
mean-field Smoluchowski equation for heterogeneous growth. However,
we shall find that his equation is incorrect since it does not conserve the
number of droplets when  the collision term is suppressed.

 Indeed, if we assume that  droplets do not coalesce, we can find the contribution
to Smoluchowski's equation due to the growth of individual droplets.
The corresponding term must conserve the number of particle, since no
new droplet is introduced in the system: actually, the equation is just
a continuity equation for the distribution function $N(s,t)$, and we find,
\begin{equation}
\partial_t N(s,t)
+\partial_s (s^{\beta} N)(s,t)=0.
\end{equation}

 If we bring coalescence  into the picture, 
we note, following Vincent \cite{vincent71}, that the rate of coalescence
of two droplets of masses $s_1$ and $s_2$ is, under the mean-field assumption,
 the time derivative of the cross-section $\propto (s_1^{1/D}+s_2^{1/D})^d$,
which is proportional to
 $(s_1^{1/D-1}ds_1/dt+s_2^{1/D-1}ds_2/dt)(s_1^{1/D}+s_2^{1/D})^{d-1}$. So that
eventually, 
\begin{eqnarray}\label{smol.hetero}
&\partial_t N(s,t)
+\partial_s (s^{\beta} N)(s,t)=\nonumber \\
&\frac{1}{2} \int_{0}^{s}  N(s_1,t)N(s-s_1,t)K(s_1,s-s_1) ds_1
\nonumber\\
&  -N(s,t)\int_{0}^{+\infty}N(s_1,t)K(s,s_1) ds_1,
\end{eqnarray}
with,
 \begin{equation}
     K(x,y)=(x^\frac{\omega}{D}+y^\frac{\omega}{D})(x^\frac{1}{D}
              +y^\frac{1}{D})^{d-1}.
     \end{equation}
Once again, redefinition of the time and mass unit was used to set 
multiplicative constants to one in the equation.

Vincent \cite{vincent71}, considered the case $d=2$, $D=3$, 
$\omega=-2$, relevant 
to the growth of epitaxial films. He derived Smoluchowski's equation for
the radius distribution $\psi(r,t)=Dr^{1-1/D}N(r^D,t)$, and found the correct
collision kernel, but his equation does not conserve the number of particles
when the kernel is set to zero, {\it i.e.} when collisions are ignored. The reason
is that he erroneously derived that the change in $\psi$ due to growth alone
was $-r^{-2}\partial_r \psi $, instead of the correct 
$-\partial_r (r^{-2}\psi)$. As a consequence, some additional
unphysical droplets are created by his equation. This might be one of the
 reasons
why Vincent had to include three and four-body coalescence events
in his Smoluchowski equation to recover correct values for the fraction
of area covered by the droplets, but the incorrect right-hand side in his
Smoluchowski equation may as well have  only minor consequences in his
approximate computation. 

An interesting case arises
 when $\omega=1+d-D$, since, as noticed by Family and 
Meakin \cite{family89}, this corresponds to the growth rate of large 
droplets due to absorption of deposited small droplets in homogeneous
growth. In this case, Smoluchowski's equation describing homogeneous 
growth differs from the one describing heterogeneous growth only by the 
injection term. 
 Exponents $\theta$ and $z$ are the same 
for both models, and numerical simulations
\cite{family89,meakindrop92} show that the  scaling function for 
large droplets 
in homogeneous growth is very similar to  the whole scaling function of 
heterogeneous growth.   

\vspace{0.2cm}
Thus, both growth and coalescence, and deposition and coalescence, are 
described in mean-field by a generalized Smoluchowski equation, with
additional terms accounting for intercollision exogenous 
growth of particles (droplets). Therefore, it is  interesting to 
perform a general
 study of this equation (with a generic kernel), and to see if its scaling 
behavior is consistent with the numerical results for droplets nucleation.

\section{Smoluchowski's equation with growth and injection}\label{sec:smol}
We consider the following generalized Smoluchowski equation,
\begin{eqnarray} \label{gensmol}
&\partial_t N(s,t) +\partial_s(s^\beta N(s,t))=
 \nonumber \\
&\frac{1}{2}\int N(s_1,t)N(s-s_1,t)K(s_1,s-s_1)\,ds_1\nonumber\\
&- N(s,t) \int N(s_1,t) K(s,s_1) \,ds_1 + I(t)\delta(s-1),
\end{eqnarray}
$K(x,y)$ being a general homogeneous kernel with exponents $\lambda$ and
$\mu$ defined as in Eq. (\ref{genscal1}).

The equation describes a set of particles or clusters which collide  with 
a mass-dependent   collision
rate $K$, and grow between collisions with (see below),
\begin{equation}
\dot{s}=s^\beta.
\end{equation}  
Besides,  some small particles (monomers) are injected 
with the injection rate $I(t)$, with the possibility that $I$ be a
functional of $N(s,t)$, as found for deposition and coalescence in Sec. 
\ref{derive}.  A discrete version of this equation without the monomer
injection term and with a {\it constant collision kernel} has been 
investigated for $0\leq \beta \leq 1$ by Krapivsky and Redner 
\cite{krapivsky96}. We shall see below that their results are independently 
recovered as special cases of our general discussion of the continuous 
equation, in the scaling regime where the
discrete structure of the equation plays no role.

When the growth and injection terms are absent, the equation reduces to
Smoluchowski's equation, and its scaling properties have been extensively 
studied \cite{vdg85prl,vdg87scal,nontriv97}, but it is in no way trivial.
Even in this case, very few analytical
solutions of Smoluchowski's equation are available. For the constant kernel
$K(x,y)=1$, an exact solution is known \cite{smoluchowski18},
 with $N(s,t)\sim 4/t^2 e^{-2s/t}$.
Other solutions concern the kernels $x+y$ 
\cite{ziff84} and $xy$ \cite{ernst84}. Despite its apparently simple
structure, Smoluchowski's equation is yet
another example of a highly nontrivial mean-field theory.  

In the following, we shall study the large time properties of the solutions
of Eq. (\ref{gensmol}), and we shall  exhibit a rich diversity 
of behaviors depending on the parameters $\beta$ (characteristic of exogenous
growth), $\lambda$ and $\mu$ (characteristic of the collision kernel).

\subsection{Gelation criterion}\label{sec:gel}
A first interesting question is the possible occurrence of  a 
gelation transition 
for such equations. Gelation corresponds to the formation of an infinite 
cluster {\em at finite time}.  Without a growth term, 
nongelling kernels  correspond to $\lambda\leq 1$ \cite{vdg85prl,vdg87scal}. 
How is this modified ?  In the absence of an infinite cluster, 
the evolution equation for the total mass in the system $M_1(t)$ is obtained
by multiplying Smoluchowski equation by $s$ and integrating over all masses,
\begin{equation}\label{massevol}
\dot{M_1}(t)= M_\beta(t)+I(t),
\end{equation} 
which is physically obvious from $\dot{s}=s^\beta$. To discuss gelation, 
we have to be more cautious. Adapting the argument for standard
Smoluchowski's equation \cite{leyvraz82,hendriks83,vdg85prl,vdg87scal},
  let us consider the mass transfer   from
 clusters of
masses $s\leq L$ towards clusters of masses $s>L$, 
\begin{equation}
J_L(t)=-\int_0^Ls\,\partial_t N(s,t)ds +\int_0^L 
\dot{s}\, N(s,t)ds + I.
\end{equation}

From Eq. (\ref{gensmol}), we get, 
\begin{eqnarray}\label{massflux}
J_L(t)&=&L^{1+\beta}N(L,t)\nonumber \\
&+&\int_0^Ldx\, x N(x,t)\int_{L-x}^{+\infty} dy\, 
K(x,y)N(y,t), 
\end{eqnarray}
where the first term is the mass flux through $s=L$ due to the growth of
individual particles, while the second term is the mass flux due to
collisions.
If there is no gelation $J_L(t)$ must vanish
 when $L\to \infty$, and  Eq. 
(\ref{massevol}) holds at any time. If there is gelation at $t=t_g$,
there is an infinite cluster, or gel, in the system for $t>t_g$, and 
$J_L(t)$ is nonvanishing for $t>t_g$. At the gel point, $J_\infty(t)=
\lim_{L\to \infty} J_L(t)$ may be 
infinite, but not for $t>t_g$. The post-gel distribution must have a slowly
decaying large $s$ tail in order that $J_\infty(t)$ be finite. If we make 
the ansatz $N(s,t\geq t_g)\sim A(t)\,s^{-\tau}$ at large $s$, 
\bleq
\begin{eqnarray}
L^{1+\beta}N(L,t) &\sim& A(t) L^{1+\beta-\tau}\\
\int_0^L\!\!dx\, x N(x,t)\int_{L-x}^{+\infty}\!\! dy\, 
K(x,y)N(y,t) &\sim& A(t)^2 L^{3+\lambda-2\tau} \int_0^1\!\!dx
 \int_{1-x}^{+\infty}\!\!
dy \,x K(x,y)(xy)^{-\tau}.
\end{eqnarray} 
\eleq
We see that if gelation occurs, $\tau$ must be equal to 
$\max(1+\beta,(3+\lambda)/2)$. In the post-gel regime, the total mass 
contained in the sol phase ({\it i.e.} finite mass clusters) must be finite, which
imposes $\tau>2$. We conclude that no gelation occurs for 
$\max(\lambda,\beta)\leq 1$. Anyway, $\beta>1$ is forbidden since it leads
to explosive growth of individual particles. If $\beta=1$, Eq. (\ref{massevol}) yields 
$M_1(t)=e^t$, the total mass growth is faster than any power of $t$,
which is the  smoking gun of the gelling-nongelling boundary.

\subsection{No injection}
For a while, we specialize to the case $I=0$, corresponding 
to growth and coalescence. We  first exhibit two exact solutions, then
we  make a complete study of the scaling solutions of the general 
equation.

\subsubsection{Exact solutions}
We can solve  Eq. (\ref{gensmol}), in the case  $K(x,y)=1$,
 for $\beta=0$ and 
$\beta=1$. To do this, we consider the Fourier-Laplace transform $Z(z,t)$
of $N(s,t)$,
\begin{equation}
Z(z,t)=\int_0^{+\infty} e^{-zs}N(s,t) \,ds, 
\end{equation}
$Z(0,t)$ being the total density  of clusters $n(t)$.
  
For $\beta=0$, the Laplace transform of Eq. (\ref{gensmol}) reads,
\begin{equation}
\partial_t Z +zZ=\frac{1}{2} Z^2- Z(0,t)Z.
\end{equation}
This equation  is easily solved for $Z$. With $n(0)=1$ and
$Z(z,0)=Z_0(z)$, we find that $Z(0,t)=n(t)=2/(t+2)$ and,
\begin{equation}
Z(z,t)=\frac{e^{-zt}}{(t+2)^2\left( \frac{1}{4Z_0(z)}-\frac{1}{2}\int_0^t 
\frac{e^{zt'}}{(t'+2)^2} dt'\right)},
\end{equation}
which leads in the scaling regime $t\to \infty$, and for a monodispersed
initial condition $N(s,t)=\delta(s-1)$, to,
\begin{equation}
N(s,t)\sim \frac{2}{t^2\ln t} e^{-\frac{s}{t\ln t}}.
\end{equation}
The total mass in the system is,
\begin{equation}
M_1(t)=-\partial_z Z(z=0,t)=1+2\ln(t+\frac{1}{2}).
\end{equation}
For $\beta=1$, the equation for $Z$ is,
\begin{equation}
\partial_t Z-z\partial_z Z= 
\frac{1}{2} Z^2- Z(0,t)Z,
\end{equation}
and once again $n(t)=Z(0,t)=2/(t+2)$. 
If we choose the variable $u=ze^t$, the equation reduces to a first order differential
equation in time, and the solution is,
\begin{equation}
Z(z,t)=\frac{2}{t+2}\frac{2Z_0(ze^t)}{\left(1-Z_0(ze^t)\right)(t+2)+2Z_0(ze^t)}.
\end{equation}
With an initial monodispersed distribution, $Z_0(z)=e^{-z}$,
$Z(z,t)$ has a pole at $z_0(t)=-e^{-t}\ln (1+2/t)$, and we can explicitly
compute $N(s,t)$,
\begin{equation}
N(s,t)=\frac{4}{(t+2)^2 e^t} \exp\left(-s e^{-t} \ln(1+2/t)\right),
\end{equation}
which leads in the large time limit to,
\begin{equation}\label{beta1ex}
N(s,t)\sim \frac{4}{t^2e^t}e^{-\frac{2s}{te^t}}.
\end{equation}

The corresponding solutions for the discrete Smoluchowski equation have
been independently derived by Krapivsky and Redner \cite{krapivsky96}, 
and coincide with the solutions above  in the scaling limit. 
This coincidence was to be expected since, in the scaling regime, the 
divergence of $S(t)$ leads to the oblivion of
the discrete structure of the equation. This provides a solid confirmation of
the surprising result that, although the growth term did not change
 the scaling
function,  still equal to  $f(x)=4e^{-x}$, the scaling is no longer of 
the form $N(s,t)\sim S(t)^{-\theta} f(s/S(t))$. Rather,    the scaling is 
$N(s,t)\sim Y(t)^{-1} f(s/S(t))$, where $Y(t)\propto S^2(t)/M_1(t)$ is not a
power of $S(t)$.
 This form enforces $n(t)\propto 1/t$, a
result that holds for any $\beta$ between 0 and 1.

A consequence of the logarithmic correction in the case $\beta=0$,
 is that in contrast  to 
what we stated for the generic case in Sec. \ref{droplets}, $x_0$ is
equal to zero. The reason is that $S(t)$ in this case grows faster than
do individual particles  in absence of collisions. Hence $s_0(t)/S(t)\propto
1/\ln t$ goes to zero. This point will be fully discussed in the
case of a general kernel under the dynamic scaling assumption.
For $\beta=1$, we see that the scenario discussed in Sec. \ref{droplets}
occurs, $S(t)\sim te^t$ corresponds to a slowly vanishing $\epsilon(t)$ in
Eq. (\ref{rengro}).

\subsubsection{Scaling theory}\label{sec:scalth}
We would like to study the scaling properties of Smoluchowski's equation.
Even though Smoluchowski's equation results from an
approximation,
its scaling behavior is usually exact above an upper 
critical dimension $d_c$, and is in many cases qualitatively correct 
even below $d_c$.

Some simple arguments may give a qualitative understanding of the different
regimes to be expected for the equation.
Indeed,  
if we suppress the  collision term
(i.e. the right hand side) in Eq. (\ref{gensmol}), we are left with a
continuity equation  describing a set of
particles which grow in time with $\dot{s}=s^\beta$, and is associated with the  mass scale $S_g(t)\propto
t^{1/(1-\beta)}$.  

Conversely, if we suppress the exogenous growth term $\partial_s(s^\beta N)$
 in the left hand side, we
are back with a standard Smoluchowski equation describing clustering 
with mass conservation. The scaling properties of this equation are
well-known \cite{vdg85prl,vdg87scal,nontriv97}. The typical mass
in the scaling regime is $S_c(t)\propto t^{1/(1-\lambda)}$, and
$\theta=2$.

Thus, when both exogenous growth and collisions are active, 
we expect to observe a 
``competition'' between the two dynamic mass scales $S_c$ and $S_g$.
 If $\beta <\lambda$, $S_g(t)\gg S_c(t)$,
and in the scaling regime we expect $S(t)\propto S_g(t)$ and
$z=1/(1-\beta)$. If $\beta>\lambda$, on the contrary, the typical mass of
particles increases essentially due to collisions, hence 
$S(t)\propto t^{1/(1-\lambda)}$ and $z=1/(1-\lambda)$. 
In the marginal case $\lambda=\beta$, 
 logarithmic corrections to $S(t)$ may be observed. In fact, we know
from the exact solution of $K=1$, $\beta=0=\lambda$ that such corrections 
actually occur.  This leads us to a slightly more general scaling assumption
than the one we made for droplets coalescence models,
\begin{equation}\label{genscal}
N(s,t) \sim Y(t)^{-1}f\left(\frac{s}{S(t)}\right).
\end{equation}
We do not assume a priori that $Y(t)\propto S(t)^{-\theta}$, since we know
from exact solutions that it is not always true.

Notice that the scaling function $f(x)$ 
is not uniquely defined by Eq. (\ref{genscal})
unless we give a precise definition of $S(t)$ and $Y(t)$.  For instance,
a widely used  definition of $S(t)$, (and the one  actually 
used in numerical simulations),  is 
\begin{equation}\label{s2/s}
S(t)=\frac{<s^2>}{<s>}.
\end{equation}
However if we know a scaling function $f_s(x)$ for given definitions of $Y$ and
$S$, any other  scaling function $f(x)$ corresponding to other definitions, 
is related to $f_s$ by 
\begin{equation}\label{kappaomega}
f(x)=\kappa f_s(\xi x),
\end{equation} $\kappa$ and $\xi$
being two constants. 

From the  picture above, it is obvious that the physical cut-off, 
i.e. the mass $s_0(t)$ below which $N(s,t)$ is strictly zero, 
scales as $S_g(t)$. This is just the translation in terms of Smoluchowski's
equation of the  discussion we had for droplet growth and coalescence.
Since $S(t)\geq s_0(t)$, either $S(t)$ and $S_g(t)$ have the same scaling and the scaling function $f(x)$
is zero below a certain argument $x_0>0$, or $S(t)\gg S_g(t)$,
$x_0$ is equal to zero, and $f$ may have  a small $x$ divergence
$f(x)\propto x^{-\tau}$, with a
polydispersity exponent $\tau\geq 0$.

The scaling of the moments of the distribution $N(s,t)$ is altered by the
existence of a polydispersity exponent. 
\begin{equation}
M_\alpha= \int_{s_0(t)}^{+\infty} s^\alpha N(s,t)\, ds \sim
\frac{S^{1+\alpha}}{Y}\int_{\frac{s_0(t)}{S(t)}}^{+\infty} x^\alpha f(x)\,dx.
\end{equation}
If there is no polydispersity exponent or if $\tau<1+\alpha$, the integral
tends to a finite limit when $t\to \infty$, and,
\begin{equation} \label{scalnorm}
M_\alpha(t) \propto \frac{S^{1+\alpha}}{Y}.
\end{equation}
If $\tau>1+\alpha$, the integral  diverges and,
\begin{equation}\label{anormscal}
M_\alpha(t) \propto \frac{S^{\tau}}{Y}s_0(t)^{1+\alpha-\tau}.
\end{equation}
Finally, if $\tau=1+\alpha$,
\begin{equation}\label{logscal}
M_\alpha(t) \propto \frac{S^{1+\alpha}}{Y}\ln\frac{S(t)}{s_0(t)}.
\end{equation}

Under the general scaling assumption, we  get the following scaling
for the different terms of Smoluchowski's equation:
\begin{equation}\label{1scal}
\partial_t N(s,t) \sim 
-\frac{1}{Y}\left(\frac{\dot{Y}}{Y}f(x)+\frac{\dot{S}}{S}xf'(x)\right) ,
\end{equation}
\begin{equation}\label{2scal}
 \partial_s(s^\beta N(s,t))\sim \frac{S^{\beta-1}}{Y} 
(x^\beta f)'(x),
\end{equation}
\begin{equation}\label{3scal}
\mbox{\rm Collision term}\sim \frac{S^{1+\lambda}}{Y^2} (...).
\end{equation}
Another important equation is Eq. (\ref{massevol}) for the 
evolution of the total mass in the system, which, in the absence
of injection, reduces to,
\begin{equation}\label{Mpoint}
\dot{M_1}=M_\beta.
\end{equation}

Now, it is possible to find the asymptotics of $M_1$, $Y$ and $S$, depending
on the values of $\lambda$ and $\beta$, under the sole scaling assumption.
In fact, though the line followed in the demonstration is quite simple,
details are rather intricate due to the multiple cases
to be examined. A full length discussion is given in Appendix
\ref{app:scal}, and results are summarized in 
Table \ref{tab:scalres}. Here, we shall  only comment 
some  interesting points.
\bleq
\begin{table}
\begin{center}
\begin{tabular}{|c|c|c|c|}
\hline
parameters &  $M_1(t)$ & $S(t)$ & $Y(t)$  \\
\hline
$1>\beta>\lambda$ & 
$\displaystyle M_1\propto S^{\beta-\lambda}$ &
$ \begin{array}{l}
    			\displaystyle S(t)\propto t^z\\
  			\displaystyle z=\frac{1}{1-\beta}
		\end{array}$ &
$\displaystyle \begin{array}{l}
 		\displaystyle Y\propto S^\theta \\
		\displaystyle  \theta=2+\lambda-\beta 
		\end{array}$	
\\[4pt]      
\hline
$1>\lambda>\beta$ &
 $M_1 \to \mbox{constant}$ & 
$\displaystyle z=\frac{1}{1-\lambda}$ & 
$\theta=2$  \\
\hline
$1>\lambda=\beta$ &
$\displaystyle M_1\propto \left\{\begin{array}{lr}
				  \ln t,& \text{if}~\mu\leq 0\\
				  (\ln t)\ln(\ln t),& \text{if}~\mu>0 
	                         \end{array}\right.$&
$\displaystyle  \begin{array}{l}
  		\displaystyle S(t)\propto (tM_1)^z\\
  		\displaystyle z=\frac{1}{1-\beta}
		\end{array} $ &
 $\displaystyle Y \propto \frac{S^2}{M_1}$ \\
\hline
$\begin{array}{c}
	\beta=1\\
	 0<\lambda<1
\end{array}$ &
 $M_1=M_1(0)e^t$ & 
$\displaystyle S(t)\propto e^\frac{t}{1-\lambda}$ &
 $Y\propto S^2e^{-t}$ \\
\hline
$\begin{array}{c}
	\beta=1\\
	 \lambda=0
\end{array}$ &
 $M_1=M_1(0)e^t$ & $S(t)\propto t e^t$ & $Y
\propto tS$ \\ 
\hline 
$\begin{array}{c}
	\lambda=1>\beta\\
	 \mu>0
\end{array}$ &
$M_1 \to \mbox{constant}$&
$\displaystyle S(t)\propto e^{b\sqrt{t}}$ &
$Y\propto S^2\sqrt{t}$ \\
\hline
$\begin{array}{c}
	\lambda=1>\beta\\
	\mu\leq 0
\end{array}$ &
$M_1 \to \mbox{constant}$ &
$\displaystyle S(t)\propto e^{bt}$ &
$\displaystyle Y\propto S^2$ \\
\hline
$\begin{array}{c}
	\lambda=\beta=1\\
	\mu\leq 0
\end{array}$ &
$M_1=M_1(0)e^t$&
$\displaystyle S(t)\propto e^{be^t}$ &
$\displaystyle Y\propto S^2e^{-t}$ \\
\hline
$\begin{array}{c}
	\lambda=\beta=1\\
	\mu > 0
\end{array}$ &
$M_1=M_1(0)e^t$&
$\displaystyle S(t)\propto e^{b\sqrt{e^t}}$ &
$\displaystyle Y\propto S^2e^{-t/2}$ \\
\hline
\end{tabular}
\caption{Results of the scaling theory}
\label{tab:scalres}
\end{center}
\end{table}
\eleq

The scaling theory is consistent with the qualitative discussion above 
based on the idea of competing dynamical scales. 
It is found, that for $\lambda<\beta<1$, $S(t)$ scales as 
$s_0(t)\sim t^{1/(1-\beta)}$, $Y(t)\propto S(t)^{-\theta}$, with
$\theta=2+\lambda-\beta$, and  the scaling function is zero below a 
finite $x_0$.  If we come back to droplets models, this $\lambda<\beta$ 
condition just corresponds  to $d<D$, and we find $\theta=1+d/D$. Hence, the
scaling results of the mean-field theory are in full agreement with the 
discussion and results  in Sec. \ref{droplets}. 

For $\lambda>\beta$, $S(t)$ scales as $t^{1/(1-\lambda)}$, the mass is 
asymptotically conserved with $\theta=2$. The scaling function may have
a polydispersity exponent, since now $S(t)\gg s_0(t)$,
 and the scaling equation is,
\begin{eqnarray}\label{scaleq}
&b\left[xf'(x)+ 2f(x)\right]=f(x)\int_0^{+\infty} f(x_1)K(x,x_1)dx_1  \nonumber\\
&- \frac{1}{2}\int_0^x f(x_1)f(x-x_1)K(x_1,x-x_1) \,dx_1,
\end{eqnarray}  
that is the same scaling equation as for standard Smoluchowski's equation
without growth of particles, which makes it possible to use all the
corresponding results or techniques \cite{vdg85prl,vdg87scal,nontriv97}.
This case corresponds to $d>D$, but as further discussed in Sec. 
\ref{heteropoly}, heterogeneous growth is always gelling in this case, and
the mean-field approximation breaks down.
 
For $\lambda=\beta$, we find that $S(t)$ is no longer a pure power law, but 
incorporates  logarithmic corrections. The total mass in the system
increases logarithmically, $M_1(t)\propto \ln t$ and $S(t)\sim
(tM_1(t))^{1/(1-\beta)}$. Once again, $S(t)\gg s_0(t)$ and the scaling
function is Eq. (\ref{scaleq}). Thus, there is a polydispersity exponent 
if the kernel has $\mu\geq 0$ (see below), and there is an addition 
$\ln(\ln t)$ correction for $\mu>0$ kernels. For heterogeneous growth,
 $\lambda=\beta$ corresponds to $d=D$, and  the mean-field theory accounts
for the qualitative difference between $d=D$ and $d<D$ observed in numerics
(see \cite{family89} and below). This point will be fully discussed 
in Sec. \ref{heteropoly}.
 This also recovers
the scaling behavior of the exact solution for $K=1$ and $\beta=0$.

 For $\beta=1$, the scaling of the exact solution $K=1$, $\beta=1$ is 
recovered. For $\lambda\geq 0$, the scaling equation is once again 
Eq. (\ref{scaleq}). Other results  in Table
\ref{tab:scalres}, show the  great diversity of scaling regimes depending on
$\beta$, $\lambda$, and  $\mu$. 

For a constant kernel, and $1>\beta>0$, we recover the result of Krapivsky
and Redner \cite{krapivsky96}. The latter authors  assumed 
that the scaling function has essentially the same shape as in the case
$\beta=0$ or the pure aggregation case. 
From our analysis, we know that this assumption  is actually 
not verified. However, it can be seen that 
the key point of their demonstration is that $f(x)$ has no small $x$
divergence, which is indeed true.

The fact that the scaling results of the $d<D$ growth and coalescence
are recovered by Smoluchowski's equation approach gives a firm basis 
to the heuristic arguments used to find {\it a posteriori} the exponents
from the obtained numerics. 
Moreover the kinetic equation approach is predictive, and
provides a synthesized classification of the aggregation models, depending
on a limited number of relevant parameters, provided that the approximation 
be justified.

\subsubsection{Polydispersity exponents} \label{sec:polydis}
    An interesting corollary of the scaling theory of generalized Smoluchowski's
equation with growth, is that in the cases $\beta\leq\lambda$ and $\beta=1$,
 the scaling equation is exactly the same as for standard 
Smoluchowski's equation, Eq. (\ref{scaleq}),
where $b$ is called the separation constant, which we set to one by
absorbing it in the scaling function (which corresponds to a redefinition of $Y$). Note that if $f$ is a solution
of Eq. (\ref{scaleq}), $c^{1+\lambda}f(cs)$ is also a solution,
 which corresponds to different possible definitions of $S(t)$ (remember
the discussion above Eq. (\ref{kappaomega})).

Thus,  all the results known for the scaling function of standard
Smoluchowski's equation also hold for the generalized equation.
For instance the scaling function $f(x)$ of the $K=1,\beta=1$ case
 can be derived 
from the exact result for standard Smoluchowski's equation with $K=1$, for
which $f_0(x)=e^{-x}$ is a
scaling function. For a given definition of $S$ and $Y$, the corresponding
scaling function for $K=1,\beta=1$  is obtained using $f_s=f_0$ in
Eq. (\ref{kappaomega}).
 If we use Eq. (\ref{s2/s}) as a definition of $S(t)$, $\xi$ is 
constrained to the value $\xi=2$ and if we define $Y(t)$ by
$M_1(t)=S^2/Y$, we get $\kappa=4$, which leads to,
\begin{equation}
N(s,t)\propto \frac{4 e^t}{S^2} e^{-2s/S},
\end{equation} 
with $S(t)\propto te^t$
in agreement with the exact result Eq. (\ref{beta1ex}).

 We can also find the exact scaling function for $\beta=1$ and $K(x,y)=x+y$
(which corresponds to $\lambda=1$). For standard Smoluchowski's equation, 
a scaling function is $f_1(x)=x^{-3/2} e^{-x}$. Thus, it is also a scaling
function for Eq. (\ref{gensmol}), and we obtain the exact result
$\tau=3/2$.

The scaling equation Eq. (\ref{scaleq}) for a general kernel was extensively
studied in the literature. 
 Van Dongen and Ernst \cite{vdg85prl,vdg87scal} showed that  the qualitative 
shape of the scaling function $f(x)$ at small $x$ depends on two 
parameters, the homogeneity degree of the kernel $\lambda$ and the exponent 
$\mu$ defined by Eq. (\ref{genscal1}).
 
For $\mu<0$, the scaling function vanishes as $\exp(-\alpha x^\mu+
o(x^\mu))$ at small $x$, and there is no polydispersity exponent.
 
 For kernels with $\mu>0$, there is
a polydispersity exponent $\tau = 1+\lambda$. For $\mu=0$, there is also
polydispersity, but with a nontrivial exponent $\tau<1+\lambda$,
\begin{equation}\label{taueq}
\tau=2-\int_0^{+\infty} x^\lambda f(x) dx.
\end{equation}
The determination of $\tau$ for $\mu=0$ was a challenge until recently,
because solving numerically
Smoluchowski's equation proved rather heavy, and often unsuccessful.

For the most
studied $\mu=0$ kernel, 
\begin{equation}\label{kdd}
K_D^d(x,y)=(x^\frac{1}{D}+y^\frac{1}{D})^d,
\end{equation}
which corresponds for instance to Brownian coalescence (in a $d+2$
dimensional space) with mass-independent
diffusion constant,
very few values of the exponent $\tau$ were known. In $d=1$,
Kang et al. \cite{kang86}, noticed that direct numerical resolution 
of Smoluchowski's equation did not reach the actual scaling regime, but 
a pseudo-asymptotic regime, with apparent scaling but wrong exponents
(some exact bound were known for $\tau$ \cite{vdg85jpa,nontriv97}).
For $D=1$ and $d\leq 1$, the scaling regime could be reached by 
Krivitsky \cite{krivitsky95}, leading to the determination of $\tau$ for 
ten values of $d$ between $0$ and  $1$. Song and Poland \cite{song}, used a power series
in time expansion method, to treat the cases ($d=1,D=2$), and ($d=2,D=3$), but 
we  showed \cite{nontriv97} that their result in the latter case 
is in contradiction with some exact bounds.

 We  recently introduced   a general
variational method for computing  accurate values of
$\tau$ at very low numerical cost \cite{nontriv97}.
This method was used to make a complete study of the polydispersity exponent
of the $K_D^d$ kernel, for a wide range of $d$ and $D$.  
We chose to start directly from the scaling (infinite time limit) equation
Eq. (\ref{scaleq}). We did not try to solve this equation for the whole
scaling function, but focused on the determination of $\tau$. The key
relations which we used are, on the one hand, integral equation 
 Eq. (\ref{taueq}) for $\tau$, and on the other hand a series of integral
equations for the moments $M_\alpha$ of $\tau$ \cite{vdg85prl}
 obtained by multiplying
Eq. (\ref{scaleq}) by $x^\alpha$ and integrating over $x$, for any value
of $\alpha>\tau-1$ (such that the integrals converge in zero),
\bleq
\begin{equation}\label{eqal}
 2(1-\alpha) \int_0^\infty \!\! x^\alpha f(x) \,dx=
 \int\!\!\int_0^\infty\!\!
 f(x)f(y)K(x,y)
\left[ x^\alpha+y^\alpha -\right.\left.(x+y)^\alpha\right]dx dy.
\end{equation}
\eleq
Combining these equations one obtains,
\begin{equation} \label{mean}
\tau=2-(1-\alpha)\frac{\int\!\!\int_0^\infty g(x,y)\,dxdy}
{\int\!\!\int_0^\infty g(x,y)A(x/y)\,dxdy},
\end{equation} 
with $g(x,y)= f(x)f(y) (x^\alpha y^\lambda+ x^\lambda y^\alpha)$
and $A(u)=(1+u^\alpha-(1+u)^\alpha)K(1,u)/(u^\alpha +u^\lambda)$.

The ratio in Eq. (\ref{mean}) is the inverse of the average of $A(x/y)$
with weight $g(x,y)$, so that computing the maximum $M_\alpha$ and
the minimum $M_\alpha$ of $A$
for various values of $\alpha$ leads to exact bounds for $\tau$ that
can be used to check numerical evaluations of $\tau$ (see \cite{nontriv97}
for details), since  Eq. (\ref{mean}) implies,
\begin{equation} \label{eqbounds}
2-(1-\alpha)/m_\alpha\leq\tau\leq 2-(1-\alpha)/M_\alpha.
\end{equation}

The idea of the variational approximation is to choose a parametered family
of variational functions, and to minimize the violation of Eq. (\ref{mean})
for a well-chosen sample of values of $\alpha$. 
The key point is the choice of the variational function. As argumented 
in \cite{nontriv97}, a natural three parameters class of functions, is:
\begin{equation}
f_v(x,\tau_0,c_1,c_2)=\left(\frac{1}{x^{\tau_0}}
+ \frac{c_1}{x^{\tau_1(\tau_0)}}
+ \frac{c_2}{x^\lambda}\right)
e^{-x}.
\end{equation}
The last term corresponds to the exact asymptotic decay at large 
$x$ of the scaling function \cite{vdg85prl,vdg87scal}, while 
$\tau_0$ is the polydispersity exponent, and $\tau_1$ is the subleading
exponent in small $x$ (its value in function of  $\tau_0$ is taken to be the
same as for the exact scaling function, see below and \cite{nontriv97}).
This class of function has the correct
 large $x$ and small $x$ asymptotic behavior expected for the scaling 
function. Besides, it contains the exact scaling functions for $K=1$
and  $K=x+y$, therefore the variational approximation  yields 
the exact result for $\tau$ in these cases (as checked in \cite{nontriv97}). 

We used as error function, measuring the violation of Eq. (\ref{mean}) for 
a set of $n$ moments $\alpha_i$,
\begin{equation}
\chi^2(f_v)= \sum_i\left(\tau_0-G_{\alpha_i}(f_v)\right)^2,
\end{equation}   
where $G_{\alpha_i}(f_v)$ is the right hand side  Eq. (\ref{mean}) for 
$\alpha=\alpha_i$ and $f=f_v$. This error function is by construction
strictly zero for the exact scaling function. For the chosen 
class of variational functions, $G_{\alpha_i}(f_v)$ can be expressed in terms
of Gamma functions and simple $1D$ integrals, which makes its numerical
computation extremely fast.

In \cite{nontriv97}, the variational method was used to perform a complete
study of the kernel $K_D^d$, for $0\leq d \leq 3$ and $d\leq D \leq 7$.
Some analytical expansions were obtained for $\tau$ in the limit $d\to 0$
, $D\to \infty$ and $d\to \infty$ with $\lambda=d/D$ constant. These
analytical
results combined with exact inequalities obtained from Eq. (\ref{mean}),
were used to check the variational results. The obtained results were in
excellent agreement with the few existing numerical values for $\tau$ 
as well as with the asymptotic expansions. Due  to its extremely
 low computational
cost of the variational approximation, compared to other methods in the
literature, and its excellent accuracy, the variational approximation seems
to be a good practical solution to  the
problem of the determination of $\tau$.

From this study of generalized Smoluchowski's equation with growth of 
particles, we see that the variational method can also be used  to find
the $\tau$ exponent for this equation, when polydispersity occurs, 
i.e. when $\lambda=\beta$ or $\beta=1$.
An interesting physical application of these results is {\it heterogeneous
growth with $d=D$}, for which  Smoluchowski's equation derived in Sec.
\ref{droplets} is in the class $\beta=\lambda$. 

For this problem,
$\beta$ is equal to $1+(\omega-1)/D$ and the kernel is,
\begin{equation}
  K(x,y)=(x^\frac{\omega}{D}+y^\frac{\omega}{D})
	(x^\frac{1}{D}+y^\frac{1}{D})^{D-1}.
\end{equation}   
This kernel is  formally similar to the one describing diffusion limited 
cluster-cluster aggregation \cite{meakin83,kolb83,meakinrev92}, but the meaning of the
parameters is different. For this kernel we have,
\begin{equation}
\mu=\left\{ \begin{array}{lr}
	0,& \mbox{if } \omega \geq 0\\
	\omega/D,& \mbox{if } \omega<0 
	\end{array}
\right. .
\end{equation}
 From the scaling theory we expect that 
$S(t)\propto (t\ln t)^{z}$ with $z=D/(1-\omega)$, and  a transition from 
a polydispersed scaling function with a nontrivial 
$\tau$ exponent, for $\omega \geq 0$, to a small $x$ vanishing scaling 
function, for $\omega<0$.

Consequently, it is interesting to determine the mean-field polydispersity
exponent $\tau$ for this kernel using the variational approximation. 
This will be done in Sec. \ref{heteropoly}, in which we 
also present comparisons
of scaling results from Smoluchowski's equation approach with direct 
numerical simulations of heterogeneous growth with $d=D$.

\subsection{Constant injection}\label{constinj}
 We now turn to the case of a constant injection rate. Interest in
aggregation models with injection was originally aroused from  applications
in chemical engineering (coagulation in stirred tank reactors) and
atmosphere sciences 
\cite{klett75,mcmurry80,white82,crump82,hendriks84,vicsek85,racz85,hayakawa87}. In
these contexts, injection was often associated with a sink term.
 The emergence of the concept of 
self-organized criticality \cite{bak87,dhar90} resulted in a renewal of
interest in aggregation models with constant injection
\cite{takayasu88,takayasu89,majumdar93}, since these systems  commonly evolve
 to a steady state asymptotic power law distribution, and therefore provide 
examples of  self-organized critical systems. This behavior is assessed by
numerical simulations and exact solutions in one dimension 
\cite{takayasu89,majumdar93}.

Hayakawa \cite{hayakawa87} studied Smoluchowski's equation with injection 
of monomer. He showed that for non gelling systems, with $\lambda<1$ 
\cite{vdg85prl},
the asymptotic steady state had a power law large
$s$ decay with an exponent $\tau=(3+\lambda)/2$.
Here we shall investigate the steady state in the presence of 
a growth term with exponent $\beta$. 
We shall suppose for convenience that 
the coagulation kernel $K(x,y)$ is equal to $x^\mu y^\nu +x^\nu y^\mu$.
 $\lambda=\mu+\nu$ is the homogeneity degree of the kernel. The results  are
however true for any kernel.

We are interested in the asymptotic steady state reached by the system at
large time. We shall see that it has a large $s$ power-law decay with
an exponent $\tau$ that we are able to compute in terms of $\lambda$ and
 $\beta$. 
To achieve this program, let us call $Z_\alpha(z,t)$ the Laplace transform
of $s^\alpha N(s,t)$ defined by:
\begin{equation}
Z_\alpha(z,t)=\int_0^{+\infty}\!\! s^\alpha N(s,t)e^{-zs}\,ds ,
\end{equation}
for which we get the following equation:
\begin{equation}
\partial_t Z_1+zZ_\beta= Z_\mu Z_\nu- Z_\mu M_\nu
-Z_\nu M_\mu +I e^{-z}.
\end{equation}

Now we consider the equation for the steady state, 
\begin{equation}
zZ_\beta^\infty= (Z_\mu^\infty-M_\mu^\infty)(Z_\nu^\infty-M_\nu^\infty) +I(e^{-z}-1).
\end{equation}
The large $s$ behavior of the steady state distribution is reflected in the
small $z$ behavior in Laplace space,
\begin{equation}
Z_\alpha^{\infty}(z)-M_\alpha^{\infty} \sim c_\alpha z^{\tau_\alpha},
\end{equation}
if $M_\alpha^\infty$ is finite. If $M_\alpha^\infty$ is infinite, 
it  does not appear
in the left hand side. Note that $M_1^\infty$ is certainly infinite
because there is constant injection of monomers and no dissipation of 
mass (at finite time). As a consequence $0<\tau_\alpha<1$ for all $\alpha<1$.
If $\tau_\alpha$ is non integer then for $s\to\infty$, 
\begin{equation}
s^\alpha N(s,t=\infty)\sim \frac{c_\alpha}{2\pi} \Gamma(1+\tau_\alpha) s^{1-\tau_\alpha}.
\end{equation}
As a consequence, $\tau-1=\tau_0=\tau_\alpha+\alpha$, and,
\begin{equation}\label{eq.smallz}
z(M_\beta+I+z^{\tau-1-\beta})\propto (z^{2\tau-2-\lambda}),
\end{equation}
hence  if $\tau-1-\beta>0$, then $1=2\tau-2-\lambda$ i.e. $\tau=(3+\lambda)/2$,
whereas if $\tau-1< \beta$, $M_\beta^\infty$ is infinite and does not appear
in the left hand side of Eq. (\ref{eq.smallz}) then  $\tau-\beta=2\tau-2-\lambda$ i.e. 
$\tau=2-\beta+\lambda$.  To summarize, we find,
\begin{equation}
\tau=\left\{\begin{array}{lr}
	(3+\lambda)/2, & \mbox{\rm if } \beta<(1+\lambda)/2 \\
	2+\lambda-\beta,  & \mbox{\rm if } \beta>(1+\lambda)/2 \\
	\end{array}
     \right. .
\end{equation}

Thus, we see that the exogenous growth term brings in a new feature: 
above a critical 
growth parameter $\beta_c=(1+\lambda)/2$, the power law exponent of the 
asymptotic state depends continuously 
on $\beta$, whereas if $\beta$ is less than $\beta_c$, the exponent is
unaffected by the growth term.    

The case $\beta=(1+\lambda)/2$ requires some additional care. Interpolation
of the two regimes above would lead to $\tau=1+\beta$, and it is possible 
to show that there is a logarithmic correction 
$N(s,+\infty)\propto 1/(s^{1+\beta} \ln s)$.

For $\beta<1+\lambda$, $\tau$ has the  value  found by Hayakawa 
\cite{hayakawa87} in the absence of exogenous growth from the same 
Laplace transform arguments. However, we find that these demonstrations in
Laplace space are not very illuminating, as they do not really 
show what happens ``physically''.  Here we would
like to point out that this exponent can be directly found from a 
simple argument  for $N(s,t)$. For convenience let us first forget 
the exogenous growth term (the argument is the same), and let us consider 
the steady state condition. 
$N_\infty(s)$ is a stationary distribution, in the sense that if we start
from $N(s,t=0)=N_\infty(s)$, then the distribution does not evolve.
 From this point of view, it becomes clear that the total mass injection rate
$I$ must exactly be compensated by the mass dissipation by collisions. Thus, 
the total mass flux due to collisions must be finite.
This just means that the steady state is at a {\it gel point}, and 
the argument 
of Sec. \ref{sec:gel} can be readily adapted to obtain $\tau=(3+\lambda)/2$.
Furthermore, as the total mass is infinite in the steady state, we do not 
have the restriction $\tau>2$, which determines the gel criterion for 
gelation at finite time.  Here, the transition occurs at infinite time,
when the  system has self-organized to the critical point of a gel 
transition. 

 If we  bring the exogenous growth term into the picture,
 we can also find the exponents
from the same argument. Now, the mass injection rate is $I+M_\beta^\infty$. If 
$M_\beta^\infty$ is finite, we still find $\tau=(3+\lambda)/2$. If $\beta\geq
(1+\lambda)/2$, $M_\beta^\infty$ is diverging with such a value of $\tau$. 
Consequently, $M_\beta$ is infinite, and the steady state condition is now,
\begin{equation}
I+\lim_{L\to\infty} \left(\int_0^L s^\beta N_\infty(s)ds-C(L)\right)=0 , 
\end{equation}
where $C(L)$ is the integral  in Eq. (\ref{massflux}).
The vanishing
of the divergence 
imposes $1+\beta-\tau=3+\lambda-2\tau$, and we recover 
$\tau=2+\lambda-\beta$.

\subsection{Constant mass injection.} \label{constmass}
Now, we would like to turn back to the initial problem of homogeneous
nucleation, and the corresponding Smoluchowski equation.  Let us remark that
the collision kernel has $\lambda=2d/D-1=2\beta-1$, just on the border line
of the two regimes for constant injection.  However, we saw in 
Sec. \ref{droplets} that the injection term of small droplets  vanishes as 
$1-\phi$. In the derivation of Smoluchowski's equation for droplets
deposition and coalescence, we found $I(t)$ from
a geometrical argument. Another way of seeing it, directly from 
the Smoluchowski equation, is to impose the
additional constraint to Eq. (\ref{gensmol}) 
that the mass injection rate be a constant (as in homogeneous nucleation),
say $\dot{M_1}=1$. 
From $\dot{M_1}=M_\beta+I$, this is equivalent to $I(t)=1-M_\beta(t)$.
For droplet deposition, and with this choice of constants,  $M_\beta(t)=\phi(t)$, and the geometrical argument 
is recovered. 

Thus, we shall now discuss the case of $\lambda=2\beta-1$ and {\it constant
mass injection} $\dot{M_1}=1$, {\it i.e.},
\begin{equation}
 I(t)=1-M_\beta(t),
\end{equation}
 for $\beta<1$. Once again, we
shall make the scaling assumption of Eq. (\ref{genscal}). As in homogeneous
nucleation, $M_1(t)\propto t$ leads to $Y(t)\propto S^2/t$. A very
interesting result is that $M_\beta$ must tend to $1$ at large time. First,
it is easily seen that $M_\beta(t)$ cannot diverge. The reason, is that
the injection rate of ``area'' into existing particles is equal to 
$\beta M_{2\beta-1}$ and is  always dominated by  $M_\beta(t)$ in the 
scaling regime.  More precisely, the evolution equation for the 
occupied area fraction $M_\beta$ is obtained from Eq. (\ref{gensmol}),
and, since collisions cannot increase $M_\beta$ ($\beta<1$), we have the 
inequality,
\begin{equation}\label{ineqbeta}
\dot{M_\beta}\leq \beta M_{2\beta-1} + 1-M_\beta(t).
\end{equation}  
Then, since $\beta<1$ implies $2\beta-1<\beta$, it is possible to show 
that for any value of a possible polydispersity exponent, we have in the
scaling regime, $(M_\beta-\beta M_{2\beta -1})\sim c M_\beta$, where $c$ is 
a strictly positive constant. In fact, $M_\beta\gg M_{2\beta-1}$ and $c=1$, if
$\tau\leq 1+\beta$ while $M_\beta\propto M_{2\beta-1}$ and
$c=1/(\tau-1-\beta)-\beta/(\tau-2\beta)$, if $\tau>1+\beta$. 
  This result, combined with
Eq. (\ref{ineqbeta}),
leads to,
\begin{equation}
\dot{M_\beta}\leq  1-cM_\beta(t), \quad c>0,
\end{equation} 
which shows that  $M_\beta$ cannot diverge. It is also clear that there is
no way that $M_\beta$ could become negative, since the smaller $M_\beta$, 
the larger $I(t)$. Therefore, if we rule out any pathological oscillatory
behavior, $M_\beta$ tends to a constant $\bar{\phi}$, that may possibly 
be zero.  
 Now, if $\bar{\phi}\neq 1$, then the injection is asymptotically constant, and,
from Sec. \ref{constinj}, 
there is a critical steady state, with $N(s)\sim 1/(s^{1+\beta}\ln s)$ at
large $s$, and  $M_\beta$ diverges, which is contradictory (besides, if
$\bar{\phi}>1$, the distribution is negative near $s=1$). 
Thus $\bar{\phi}=1$, which is a nontrivial result, and was a not obviously 
fulfilled  necessary condition for our mean-field approach to correctly 
describe droplets deposition and coalescence. 

Now, let us discuss the scaling properties of the equation. 
Since the injection term is vanishing, we expect
the scaling equation (which describes large clusters) to be the same as
in the case without injection. However,  the fact that the cut-off $s_0$ 
is constant, and therefore negligible compared to $S(t)$ selects a solution
different from the one obtained without injection.

To be more precise, we know that $Y\sim S^2/t$ and that $M_\beta$ has a
finite limit. If we assume that there is polydispersity with 
$\tau\geq 1+\beta$, those two conditions lead to $S(t)\gg t^{1/(1-\beta)}$,
and we find that the scaling equation is once again Eq. (\ref{scaleq}),
which yields $\tau\leq1+\lambda<1+\beta$, in contradiction with our assumption.
Thus, $\tau<1+\beta$, and $M_\beta\propto tS(t)^{\beta-1}$, leading 
to,
\begin{equation}
 \theta=1+\beta, \quad z=\frac{1}{1-\beta},
\end{equation} 
which corresponds to the results previously obtained 
for droplets deposition and coalescence (with $\beta=d/D$). 
The scaling equation is Eq.  (\ref{bimod}), with positive $a$ and $b$. 
This equation is nonlinear and is likely to admit several classes of
solutions. We have seen that when there is no injection, a solution 
is selected which vanishes below a finite $x_0>0$. However, in presence
of injection the scaling function has no lower cut-off ($x_0=0$), and we
can have a polydispersity exponent. 

To investigate the small $x$ behavior of $f$, we  introduce the auxiliary function
$\varphi(x)=x^{\beta-1}f(x)$, which leads to a new scaling equation,
\bleq
\begin{eqnarray}
2s^{1-\beta}\varphi(s)+s^{2-\beta}\varphi'(s)- s\varphi'(s)-\varphi(s)&=&
\varphi(s) \int_\varepsilon^{+\infty}
\varphi(s_1)\tilde{K}(s,s_1)ds_1\nonumber\\
& &- \frac{1}{2} \int_\varepsilon^{s-\varepsilon}
\varphi(s_1)\varphi(s-s_1)\tilde{K}(s_1,s-s_1)ds_1,
\end{eqnarray}
\eleq
\noindent 
where $\tilde{K}(x,y)=x^{1-\beta}y^{1-\beta}K(x,y)$ ($\epsilon$ is included 
to regularize the collision terms which are separately diverging in the 
$\varepsilon\to 0$ limit \cite{vdg85prl}). 
We remark that the most diverging term for $x\to 0$ in the left hand side is
$-s\varphi'(s)-\varphi(s)$, so that, as far as 
 the determination of the asymptotic
behavior $\varphi(s)\propto s^{-\tau'}$ is concerned, we can straightforwardly generalize the results of van Dongen and
Ernst \cite{vdg85prl,vdg87scal}. The kernel $\tilde{K}$ has the 
homogeneity $\tilde{\lambda}=2+\lambda-2\beta$, and
$\tilde{\mu}=1+\mu-\beta$, and we find that the function vanishes at small
$x$ for $\tilde{\mu}<0$, that $\tau'=1+\tilde{\lambda}$ for $\tilde{\mu}>0$,
and, $\tau'$ is non trivial, with,
\begin{equation}
\tau'= 1+\int_0^{+\infty}\!\! \varphi(x)x^{\tilde{\lambda}}\,dx < 1+\tilde{\lambda},
\end{equation}
for $\tilde{\mu}=0$.

Therefore, for $\mu<\beta-1$, we have no polydispersity, while for 
$\mu>\beta-1$, we have $\tau=2+\lambda-\beta$, and  for $\mu=\beta-1$,
\begin{equation}
\tau=\beta+\int_0^{+\infty}\!\! f(x)x^{\lambda+1-\beta}\,dx < 2+\lambda-\beta.
\end{equation} 

Now, for $\lambda=2\beta-1$, we find that $\tau=1+\beta$, 
if $\mu>\beta-1$, while $\tau$ is nontrivial and strictly less than
$1+\beta$ for $\mu=\beta-1$. Hence, for $\mu\leq \beta-1$, the scaling theory
is consistent, while for $\mu>\beta-1$ there is a contradiction with 
$\tau<1+\beta$. The latter case precisely corresponds to droplets deposition
(see Eq. \ref{depkern}), and we cannot conclude.   However a consistent
scaling with a nontrivial polydispersity exponent 
could be obtained  if we include  pair correlations into  the collision
kernel and if the resulting kernel has $\mu=\beta-1$. 

This scenario 
is supported by an early numerical work of Tanaka \cite{tanaka75}. Tanaka
solved a set of coupled differential equations describing growth and 
coalescence with renucleation for $d=2$, $D=3$.
 Dynamical pair correlations due to
excluded volume were included in an approximate form. It seems quite
clear that his set of equation becomes equivalent to a Smoluchowski equation
very similar to ours in the large time limit, but with a collision kernel
modified by correlations, and Tanaka found a  bimodal droplets mass
distribution with a nontrivial 
polydispersity exponent, in  agreement with the results described in Sec. 
\ref{derive}. It would be interesting to try to determine the kernel from
his equation, although this seems quite difficult.

\section{Heterogeneous growth with polydispersity}\label{heteropoly}
In Sec. \ref{sec:smol}, we found from a mean-field approach that 
the kinetics of heterogeneous growth with $d=D$ (for instance, discs on 
a plane, or spheres in 3D), should be qualitatively different from its 
counterpart with $d<D$. From the scaling theory of generalized
Smoluchowski's equation, we found that there should be a transition from
a monodispersed scaling function for $\omega<0$, to a polydispersed 
function with a nontrivial polydispersity exponent $\tau$ for $\omega\geq
0$.  This mean-field result is actually very interesting,
 since it corroborates numerical simulations performed by Family and Meakin
\cite{family89,meakindrop92}, who found 
that polydispersity occurs for $d=D=2$ and $\omega=0.5$. 

Thus, our Smoluchowski equation approach sheds some new light on
heterogeneous growth with $d=D$, which was not much studied 
due to the fact that interest was primarily focused on $d=2,D=3$ relevant 
to breath figures, and also to the fact that numerical simulations are much 
more difficult 
in this case (see below). In this section, we  first fully discuss
what should be expected from the mean-field theory, and  compute the
polydispersity exponents for $\omega\geq 0$. Then we present some numerical 
simulations in $d=2$ and discuss the relevance of the mean-field theory.

\subsection{Mean-field theory}
In Sec. \ref{droplets}, it was found that the collision kernel corresponding
to heterogeneous growth with $d=D$ was,
\begin{equation}
K(x,y)=(x^\frac{\omega}{D}+y^\frac{\omega}{D})
	(x^\frac{1}{D}+y^\frac{1}{D})^{D-1},
\end{equation}
with $\lambda=1+(\omega-1)/D=\beta$. 
The corresponding generalized Smoluchowski equation was found to be
nongelling for $\omega\leq 1$, and in the following we shall take
$\omega<1$.

The reason why growth with $d=D$ is different in mean-field from $d<D$, is 
rather subtle. As discussed in Sec. \ref{sec:smol},
the ``competing'' dynamical mass scales corresponding to exogenous growth and 
growth by collision, respectively $S_g(t)$ and $S_c(t)$ are of the same
order at large times for $d=D$, 
which leads to a  marginal enhancement of the growth of the typical mass
$S(t)$, and to logarithmic  mass growth,
\begin{equation}
S(t)\propto (t\ln t)^\frac{D}{1-\omega}, \quad M_1(t) \sim \ln t.
\end{equation}
This implies that the cut-off $x_0=\lim s_0(t)/S(t)$ 
in the scaling function is zero,
in contrast to the $d<D$ case for which $x_0>0$, and the scaling equation
is the same as for Smoluchowski's equation without growth. 

For $\omega\geq 0$, we have $\mu=0$, and consequently there is a nontrivial
polydispersity exponent $\tau$. We can use the methods discussed in
Sec. \ref{sec:polydis} to study $\tau$.  

As a preliminary remark, let us show that when $D\geq 2$, the exponent
$\tau$ is bigger than one. Let us assume that $\tau<1$. Since the scaling
function is integrable in zero,  we can write
Eq. (\ref{eqal}) with $\alpha=0$,
\begin{equation}\label{al=0}
2\int_0^{+\infty} \!\! f(x)dx= \int\!\!\int_0^{+\infty}\!\! f(x)f(y)K(x,y)
\,dx\, dy
\end{equation}
From the inequality,
\begin{equation}
K(x,y)=(x^\frac{\omega}{D}+y^\frac{\omega}{D})
(x^\frac{1}{D}+y^\frac{1}{D})^{D-1}\geq x^\lambda+y^\lambda
\end{equation}
for $D\geq 2$, we see that Eq. (\ref{al=0})  leads to,
\begin{equation}
\int_0^{+\infty} \!\! f(x)dx \geq \int_0^{+\infty}\!\! f(x)\,dx 
\int_0^{+\infty}\!\! y^\lambda f(y)
\, dy
\end{equation}
which, combined with Eq. (\ref{taueq}), implies $\tau\geq 1$ in contradiction
with our assumption.

	With the method briefly discussed in Sec. 
(\ref{sec:polydis}), it is easy to obtain exact bounds for $\tau$ 
to control the results of the variational approximation we shall
use.  As a concrete example, let us determine such bounds
for $D=2$ and $\omega=0.5$.  Since 
$\tau<1+\lambda$ (here, $\lambda=0.75$), Eq. (\ref{mean}) holds for
$\alpha=\lambda$, for which we can 
 numerically compute the minimum and maximum of $A$. From Eq. 
(\ref{eqbounds}), this leads to 
the inequality $1.5 \leq \tau \leq 1.607175$. Thus, Eq. (\ref{eqbounds})
holds for $0.607175<\alpha\leq \lambda$, and we can compute new bounds
for each $\alpha$ in this interval, and find the tightest bounds.
 The upper bound obtained for $\alpha=\lambda$ cannot be improved
since $A(0)=1$ for $\alpha<\lambda$, hence $2-(1-\alpha)/M_\alpha \geq
1+\alpha$, but we obtain a better lower bound of $1.54$ for $\alpha=0.68$.
Table \ref{tab:ineq} presents such exact bounds for $D=2$.

For $D=1$, the kernel is equal to $x^{\omega}+y^{\omega}$,
 corresponding to the kernel $K_{1/\omega}^1$ with the notations of 
Eq. (\ref{kdd}), which was extensively studied in
\cite{nontriv97}. The exponent $\tau$ is bigger than $1$ for any $\omega>0$
while $\tau=0$ when $\omega=0$. Since $1\leq \tau <1+\omega$ for $\omega>0$,
we see that $\tau\to 1$ when $\omega\to 0$, hence $\tau$ has a discontinuity
at $\omega=0$.  

\begin{figure}
\begin{center}
\epsfig{figure=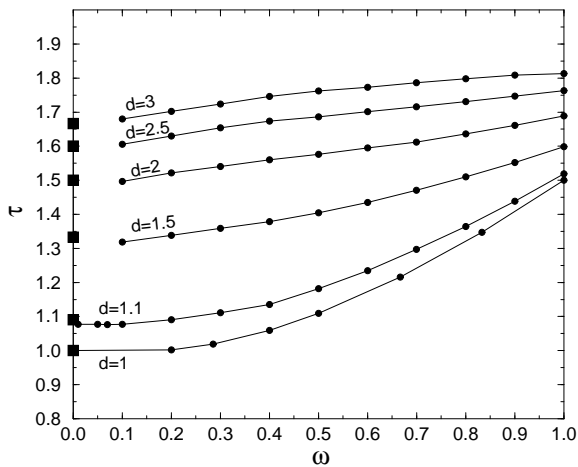,width=\linewidth}
\caption{The exponent $\tau$ for the kernel 
$(x^{\omega/D}+y^{\omega/D})^{D-1}$ was computed with the variational
approximation  for $\omega>0$ and $1\leq D\leq 3$. The theoretical $\omega
\to 0$ limit of $\tau$, $\tau_0=2-1/D$ is plotted on the Y-axis (squares).
}  
\label{fig:tauvar}
\end{center}
\end{figure}

\begin{figure}
\begin{center}
\epsfig{figure=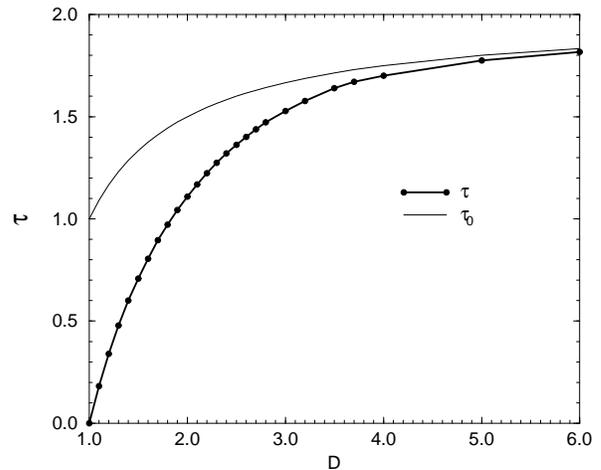,width=\linewidth}
\caption{Variational approximation for $\tau$ when $\omega=0$, compared to
 its $\omega\to 0^+$ limit $\tau_0=2-1/D$. Both $\tau$ and $\tau_0$ tend 
to $2$ when $D\to \infty$.}
\label{fig:disc}
\end{center}
\end{figure}

 The variational approximation was used to study the behavior of $\tau$.
 The set of moments was chosen as discussed in \cite{nontriv97}.
Results are shown on Fig. \ref{fig:tauvar} for different values of $D$ and
$\omega>0$, while Fig. \ref{fig:disc} shows the values of $\tau$ for $\omega=0$.
It appears that $\tau$    has a
discontinuity at $\omega=0$  not only for $D=1$, but for $D>1$ as well: when $\omega\to 0^+$, $\tau$ appears to 
have a limit $\tau_0$  bigger than its value at $\omega=0$. Thus, the
discontinuity which was rigorously shown to exist for $D=1$, pertains
for $D\geq 1$. It is difficult to accurately extract the value of $\tau_0$
since the variational algorithm appears to be less accurate for small values
of $\omega$ (for $\omega$ typically less than $0.1$). However, $\tau_0$ 
seems to be close to $2-1/D$, which is the value of $1+\lambda$ at
$\omega=0$. Actually, a heuristic argument, inspired from the discussion
for the $K_D^d$ kernel in the large $D$ ($d>1$) limit, yields
$\tau_0=2-1/D$.

  Let $f_0(x)$ be the exact scaling function for $\omega=0$. From 
Eq. (\ref{taueq}), we get,
\begin{equation}
\tau_0 = \tau_{\omega=0} + \lim_{\omega \to 0^+}\int_0^{+\infty} \!\!
(f(x)-f_0(x))x^{1+\frac{\omega-1}{D}} dx
\end{equation}
and the limit in the right hand side of the latter equation must be 
strictly positive, although $(f(x)-f_0(x))\to 0$ for any $x>0$.
How can this occur ? 
Since $\tau>\tau_{\omega=0}$ (for small $\omega$), $(f(x)-f_0(x))\sim c/x^\tau$
when $x\to 0$, and $c$ must vanish when $\omega \to 0$. 
Thus, the integral has an integrable singularity $cx^{1+(\omega -1)/D-\tau}$.
If $\tau_0<2-1/D$ (we know that $\tau_0\leq 2-1/D$ from $\tau<1+\lambda$), the contribution of the singularity is wiped out by the vanishing of $c$,
whereas,
if $\tau_0=2-1/D$, the integral is equivalent to $c/(\tau_0+\omega/D-\tau)$,
and it has the finite limit $\tau_0-\tau_{\omega=0}$ provided that $c$ vanishes as 
$(\tau_0-\tau_{\omega=0})(\tau_0+\omega/D-\tau)$. 
 
Figure \ref{fig:disc} plots the value of $\tau$ and $\tau_0=2-1/D$
 for $\omega=0$ and $1\leq D \leq 6$.  Both $\tau$ and $\tau_0$ have the
limit $2$ when $D\to \infty$ which implies that the discontinuity in
$\omega=0$ vanishes at large $D$, as can be seen on the figure. 
The reason why $\tau \to 2$ is that when $D\to \infty$, 
\begin{equation}
K(x,y) = 2^D\left[(xy)^\frac{1}{2} +O(1/D)\right]
\end{equation} 
and therefore the $D\to\infty$ limit of $\tilde{f}=2^Df$ is solution
of Eq. (\ref{smolscal}) with the kernel $(xy)^\frac{1}{2}$  which
is a $\mu>0$ kernel with exponent $\tau=2$ (the same trick was used 
in \cite{nontriv97} to study the $d\to\infty$, $d=\lambda D$ limit for the
$K_D^d$ kernel).  	  

For $\omega<0$, we have $\mu=\omega/D<0$, and using the results of van
Dongen and Ernst \cite{vdg87scal}, we  have, for $x\to 0$,
\begin{equation}
f(x)\sim  B(\omega)x^{-\gamma(\omega)}
\exp\left(\frac{D}{b(\omega)\omega}x^{\frac{\omega}{D}} 
{\textstyle \int_0^{+\infty} x^{1-1/D} f(x) dx}\right)  
\end{equation}
where, $B$, $\gamma$ and $b=\lim \dot{S}S^{-2-\lambda} Y$ 
are $\omega$ dependent constants. These
constants also depend on the definitions of $Y(t)$ and $S(t)$, but $\gamma
\to 2$ when $\mu \to 0$. For a given
definition, say 
\begin{equation}
S(t) = <s^2>/<s>, \quad Y(t)= S^2/M_1,
\end{equation}
van Dongen and Ernst showed that the scaling function crosses over to the 
$\mu=0$ (polydispersed) case when $\omega\to 0$, since the small $x$ 
asymptotics tends to $B(0)x^{-\tau}$, where $\tau=b^{-1}(2-\int_0^{+\infty}
x^{1-1/D}f(x)dx)$ is precisely the $\omega=0$ polydispersity exponent (we had
set $b=1$ in Eq. (\ref{taueq})). 

Consequently, we should observe this cross-over 
in numerical simulation. Moreover,
for small, but finite $\omega$, the critical $x_c$ below which $f(x)$  is
significantly departing from the power law corresponds to $\mu \ln x_c$ of
order one. Thus, it is reasonable to  expect  a scaling behavior when
$\omega\to 0^-$,
\begin{equation}\label{crossover}
f(x,\omega)=x_c(\omega)^{-\tau} g(x/x_c(\omega)),
\end{equation}
with $x_c(\omega) = \exp( - c/\omega +o(1/\omega))$,
$g(y) \to 0$ at small $y$, and $g(y)\propto
y^{-\tau}$ at large $y$.

\begin{table}
\begin{center}
\begin{tabular}{|l|c|r|}
\hline
$\omega$ & $\tau_m$ & $\tau_M$ \\
\hline
0.02 & 1.020 & 1.510 \\
0.2 & 1.339 & 1.588 \\
0.3 & 1.472 & 1.594 \\
0.4 & 1.514 & 1.601 \\
0.5 & 1.540 & 1.608 \\
0.6 & 1.572 & 1.614 \\
0.8 & 1.623 & 1.800 \\
0.9 & 1.633 & 1.900 \\
\hline
\end{tabular}
\caption{Exact upper and lower bounds}
\label{tab:ineq}
\end{center}
\end{table}

\subsection{Numerical simulations}
 Family and Meakin \cite{family89} and 
Meakin \cite{meakindrop92}  found from simulations  that there was a
polydispersity exponent for  $d=D=2$ and 
$\omega=1/2$, and interpreted  the qualitative difference with the $d<D$ regime
as a gelling boundary effect.

 Indeed, for the actual heterogeneous growth
model, it seems obvious
that there is gelation for all $d>D$, while the mean-field theory 
finds that there is no gelation for $\lambda\geq 1$, 
i.e. for $d\leq D+1-\omega$. This tends to show that the mean-field theory
fails at least for $d>D$. Numerical simulations performed for $d=2$,
$\omega=-1$, and $D=1.7>d+\omega-1$, show gelation at a finite time 
$t_g\approx 0.12$, which was seen to be nearly unaffected
 when doubling the mass of the sample, keeping the same initial density and
mass of the droplets, or reducing the time step 
by a factor $2$, and thus seems to be well-defined in the continuous time 
and thermodynamic limit, corresponding to a genuine gelation transition.
This is a  confirmation of the naive gelation criterion $d>D$. 
Actually, it is clear that the mean-field theory must break-down also 
for $d=D$. The reason is that the mean-field $M_1(t)$ diverges, while
the actual $M_1(t)$ cannot diverge from a geometric constraint which, of
course, is absent in the mean-field theory.  Indeed, for $d=D$, $M_1(t)$ 
is also proportional to the occupied area fraction, and is therefore
bounded. This means that at large times, strong density-density 
correlations play an important role, and are not taken 
into account in the mean-field theory.
  
In fact, the mean-field results may be qualitatively correct in an
intermediate regime between small times, and the asymptotic non mean-field 
scaling regime, and the scaling function may have a behavior
 qualitatively close from what expected for the mean-field.  

\begin{figure}
\begin{center}
\epsfig{figure=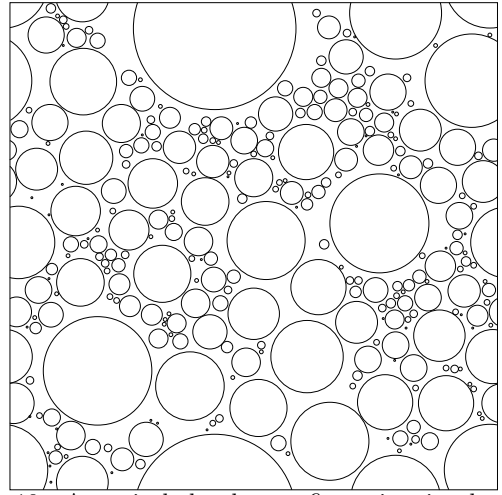,width=0.75\linewidth}
\caption{A  typical droplet configuration in the scaling regime of growth
and coalescence with $d=D=2$, obtained here for $\omega=-1$, from $262144$ 
droplets of radius $0.75$ in the initial condition on a $1024\times1024$
lattice. The picture represents the whole system (with periodic boundary conditions) at
$t=5.0$ ($S=7693.9$). The number of droplets has dropped to $287$.}
\label{om1conf}
\end{center}
\end{figure}

To check this, we performed 
simulations in $d=2$, for various values of the growth exponent $\omega$.
In one step of simulation, all the droplets radii were increased of an 
amount $\delta r=r^\omega \delta t$, then collisions were looked for and resolved.
In most of the simulations, the time increase $\delta t$ was equal to
$0.005$. It was chosen small enough such that further reduction would not
lead to significant modification of the results. As can be intuitively 
understood, the number of droplets decreases much faster for $d=D$
than for $D>d$. In the scaling regime, the number of droplets was cut by a
factor of more than $1000$, and we were obliged to start from a huge number
of droplets (about $2.5 \times 10^5$) and to perform a large number of
simulations to obtain acceptable statistics, without being able to reach
very large times. Fig. \ref{om1conf} shows a configuration obtained 
at $t=5.0$ for $\omega=-1$ from an initial configuration of $1024^2$
droplets. It is striking  that the distribution of  
droplets masses looks much broader than for $D=3$ (see Fig. \ref{fig:hetero3D}). 

\begin{figure}
\begin{center}
\epsfig{figure=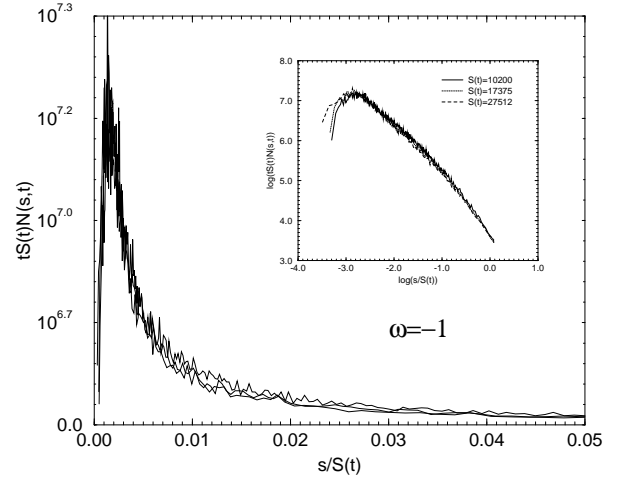,width=\linewidth}
\caption{Scaled mass distributions at three stages in the simulation of
heterogeneous growth with $\omega=-1$. These results were obtained from 96
simulations. In each simulation, $262144$  
droplets of radius $0.75$ were initially randomly placed on a $1024\times
1024$ lattice (without overlap). }
\label{om1fig}
\end{center}
\end{figure}

The scaling form Eq. (\ref{genscal}) was used with $Y=tS^{1+\beta}$
 to obtain convincing data collapse,  as shown  on Fig. \ref{om1fig}.

\begin{figure}
\begin{center}
\epsfig{figure=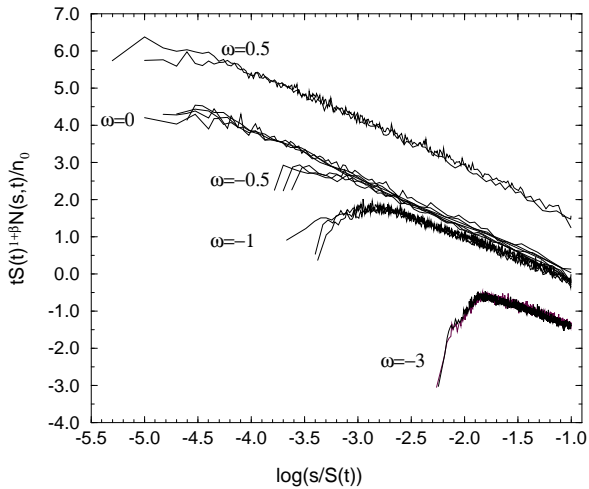,width=\linewidth}
\caption{Small $x=s/S$ behavior of the scaled mass distributions 
obtained in numerical simulations for different values of the growth
exponent $\omega$.}
\label{cross}
\end{center}
\end{figure}

Fig. \ref{cross} plots the scaling functions for various values of
$\omega$.  The results are consistent with a
transition from a small $x$ diverging 
scaling function, for $\omega\geq 0$, to 
a small $x$ vanishing scaling function for negative value of $\omega$. 
   For the considered values of negative 
$\omega$, the scaling function, as visible on Fig. \ref{om1fig}, although vanishing when $x\to 0$ is quite 
broad, with a maximum at a value $x$ significantly smaller than $1$. When
$\omega \to 0^-$, we observe a cross-over to the $\omega=0$ power law, and
the position of the maximum of the scaling
function   rapidly tends  to zero when $\omega\to 0$, consistently with the 
discussion around Eq. (\ref{crossover}).  Moreover, the exponent extracted 
from the numerics is about $1.2$, which compares well with $\tau=1.108$ from
mean-field. However, 
the $\tau$ exponents for $\omega=0$ and $\omega=0.5$,
do not seem to be significantly different, in contrast with the quite large 
discontinuity in mean-field.

\begin{figure}
\begin{center}
\epsfig{figure=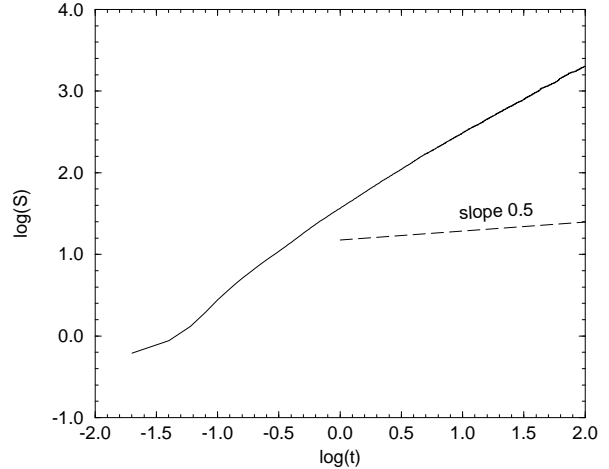,width=\linewidth}
\caption{For $\omega=-3$, $S(t)$ grows much faster than $t^{D/(1-\omega)}$
(slope $0.5$) even in the scaling regime ($t>50$) corresponding, a
behavior which may be related to logarithmic enhancement of $S$ in mean-field.}
\label{fig:s}
\end{center}
\end{figure}

As far as $S(t)$ is concerned, it is difficult to be positive due to
strong numerical limitations. Figure \ref{fig:s} shows the evolution 
of $S(t)$ for $\omega=-3$. The excellent data collapse shown in Fig. 
\ref{cross} corresponds to $t=65.0$ and $t=77.5$. At these times, 
$S(t)$ is seen to grow much faster than $t^{D/(1-\omega)}=\sqrt{t}$, which
is consistent with the mean-field logarithmic enhancement. This may explain
why the mean-field description is qualitatively correct in the 
scaling regime observed in simulations. A strongly non mean-field
scaling with properties closer to the $d<D$ case may be observed at times
unreachable to our simulations.

\section{conclusion}	
In this article, we have extensively studied a generalized   Smoluchowski
 equation corresponding  
to aggregation processes
for which  particles (or clusters) grow between collisions, with 
$\dot{s}=As^\beta$, and small particles (monomers) are injected. 

 A physical
motivation for this work is droplets nucleation and we have directly derived 
generalized Smoluchowski  equation and found the collision kernel for
two models, respectively describing homogeneous and heterogeneous
nucleation. 

For a generic kernel, with parameters $\lambda$, $\mu$, we have shown
that the gelation criterion was $\max(\lambda,\beta)>1$. We have devoted
much time to the study of the equation {\it without injection}, for which we have
provided two exact solutions. The scaling properties for a generic kernel
are seen to be strongly affected by the exogenous growth term, and depend
on $\beta$, $\lambda$, and $\mu$. For $\lambda>\beta$, however, 
the scaling is the
same as for the standard equation. For the interesting case $\lambda=\beta$,
the behavior of the typical mass $S(t)$ is modified, but the scaling
function is unchanged. For $\lambda<\beta$, the scaling function is 
qualitatively different, and vanishes at a finite $x_0>0$. 

We have also studied the case of a {\it constant injection rate} of
monomers.
 The distribution reaches an asymptotic steady state with a power law tail 
$N_\infty(s)\propto s^{-\tau}$, and 
we find that $\tau$ depends on $\beta$ and $\lambda$.
We have shown that  this steady state can be seen as a 
sol-gel critical state, breaking the mass conservation by collisions, which
we think clarifies our physical understanding of these power law asymptotic 
states.

We have payed special attention to the case of a {\it constant mass  injection rate}
and $\lambda=2\beta-1$,  related to homogeneous nucleation. 
This corresponds to a time-dependent,
self-consistent injection rate $I(t)\propto c-M_\beta(t)$.
 We have shown that $I(t)$
vanishes at large time, in agreement with the droplets deposition and
coalescence model. For droplets deposition, $M_\beta$ is proportional to 
the surface coverage, and the vanishing of the injection rate  corresponds
to  the saturation of the coverage to $1$. Thus, our self-consistent 
Smoluchowski's equation recovers a merely geometrical constraint, which is
quite nontrivial. 

As far as scaling is concerned, we have found consistent
results for $\mu\leq \beta-1$, with nontrivial polydispersity exponents
for $\mu=\beta-1$, recovering $\theta=1+\beta$ and $z=1/(1-\beta)$ as for
the droplets deposition and coalescence model. However, for $\mu>\beta-1$
kernels, we could not find a consistent scaling, and there might be no
scaling solution with a constant mass injection rate.  The mean-field kernel
for droplets deposition and coalescence has $\mu>\beta-1$, and we  argued 
that taking into account  excluded volume pair correlations may be essential 
to obtain a consistent description by a kinetic equation,  including
a nontrivial polydispersity exponent if $\mu$ is switched to $\beta-1$. It
is quite difficult however to study these correlations either numerically or
analytically.

Finally, we have applied these results to droplets growth and coalescence with
$d=D$. We have shown that Smoluchowski's approach accounts for 
the qualitative difference in the scaling function with the $d<D$ case. 
We have computed nontrivial polydispersity exponents occurring 
for $\omega\geq 0$, and 
described the cross-over from monodispersity to polydispersity occurring 
for $\omega\to 0^-$. We have compared these theoretical results with numerical 
simulations, with good agreement, despite mean-field limitations which 
we discussed. 

 As a  conclusion, we would like to point 
out that  one of the main reasons why people have become
increasingly interested in breath figures is that it is an example
of a {\it geometrically constrained} growth process, where diffusion plays a
minor role, in contrast with diffusion limited cluster-cluster aggregation
 \cite{meakin83,kolb83,meakinrev92} or
Brownian coalescence of droplets \cite{meakindyn90}.
Therefore, one could doubt that neglecting density-density correlations may 
have no dramatic consequences. Indeed, for homogeneous nucleation, we have
seen that pair correlations may be crucial to find a correct scaling
function, and we also found an infinite upper critical dimension for 
heterogeneous $d=D$ nucleation.
However, we have shown that Smoluchowski's equation in an extended form
could  be  successfully used
to  describe heterogeneous growth, and also gives very interesting 
insights into homogeneous growth, which was not a priori obvious.

\vspace{0.5cm}
We are very grateful to P.L. Krapivsky for helpful correspondence.
\appendix
\section{Scaling theory without injection}\label{app:scal}
In this appendix, we shall give a detailed demonstration of the scaling
results given in Sec. \ref{sec:scalth} for generalized Smoluchowski's
equation without injection of monomers. We make the assumption that starting
from a monodispersed distribution of droplets,
the late time solution of Eq. (\ref{gensmol}) has the scaling form 
of Eq. (\ref{genscal}).  Note that,
 although  it is quite clear from its homogeneity behavior that   
Smoluchowski's equation admits scaling solutions,
it has never been mathematically proved that these solutions are approached
at large times, except in the few cases for which we can obtain the exact
solution. However, scaling is commonly observed experimentally
 and  numerically for aggregation models, as well as in numerical
solutions of Smoluchowski's equation \cite{krivitsky95} (when possible),
 making this
assumption very reasonable. 

Another important remark is that Smoluchowski's
equation may generally speaking admit different classes of scaling solutions, 
with different asymptotic behavior, and that one peculiar class will be 
selected from
the initial condition. This must be remembered in the discussion. 
In our demonstration, we shall always use the fact that $S(t)$ cannot be
negligible compared to $s_0(t)\sim t^{1/{(1-\beta)}}$, the lower cut-off, 
from the discussion in Sec. \ref{sec:scalth}. We shall also implicitly
assume that $Y(t)$, $S(t)$ and all the moments of $N(s,t)$ are
asymptotically regular convex or concave functions.

\subsection{Within the nongelling domain}

To start with, let us study the case $\lambda<1$ and $\beta<1$,
i.e. nongelling systems
which are not on the gelling boundary. 
Our discussion is based on the fact that either $S(t)\gg s_0(t)$ 
or $S(t)\propto s_0(t)$. We shall study the implications of both 
possibilities.
\vspace{0.5cm}

\subsubsection{Case $S\gg s_0$} 

Let us assume that $S(t)\gg s_0(t)$, or equivalently that 
$S^{\beta-1}\ll \dot{S}/{S}$. We see that the growth term (\ref{2scal})
is much smaller than (\ref{1scal})   in the
scaling limit, and the scaling equation  is,
\begin{eqnarray}\label{smolscal}
&bxf'(x)+ a f(x)=f(x)\int_0^{+\infty} f(x_1)K(x,x_1)dx_1  \nonumber\\
&- \frac{1}{2}\int_0^x f(x_1)f(x-x_1)K(x_1,x-x_1) \,dx_1
\end{eqnarray}
where $b=\lim \,\dot{S}S^{-2-\lambda}Y$ and $a=\lim \,\dot{Y}
S^{-1-\lambda}$ are positive, and possibly zero or infinite.
 
For finite $a$ and $b$, this equation is very close from 
standard Smoluchowski equation Eq. (\ref{smoleq}) with the same kernel $K$,
which corresponds to $a=2b$ and $0<a<+\infty$  and was well studied in the literature \cite{vdg85prl,vdg87scal,nontriv97}. 
The polydispersity exponent $\tau$, if any, has the upper bound 
$\tau \leq 1+\lambda<2$ (read Sec. \ref{sec:polydis}). As a consequence, from Eq. (\ref{scalnorm})
we have,
\begin{equation}\label{Mscalnorm}   
M_1(t)\propto S^2/Y
\end{equation}
From Eq. (\ref{Mpoint}), we see that $M_1(t)$ is nondecreasing. Thus, either
$M_1(t)$ tends to a finite limit, or it goes to infinity.
\vspace{0.2cm}

{\it (i)} If $M_1(t)$ {\it tends to a finite limit}, then, necessarily from Eq. 
(\ref{Mscalnorm}), 
$Y\propto S^2$. The scaling of (\ref{1scal}) with 
(\ref{3scal}) requires $\dot{S}\propto S^{\lambda}$, hence,
\begin{equation}
S(t) \propto t^{\frac{1}{1-\lambda}}
\end{equation}
To be consistent with our assumption that $S(t)\gg t^{1/(1-\beta)}$, 
we must have $\lambda>\beta$. Besides, from   Eq. (\ref{Mpoint}), a necessary condition   
for  $M_1(t)$ to have a large $t$ finite limit
is that $M_\beta(t)$ must be an integrable function.
If $\tau<1+\beta$, $M_\beta$ scales as $S^{1+\beta}/Y$, i.e.
$M_\beta \propto t^{(\beta -1)/(1-\lambda)}$ and is integrable 
since $\lambda>\beta$. 
 
If $\tau>1+\beta$, we have, from Eq. (\ref{anormscal}), 
\begin{equation}\label{mbeta}
M_\beta \propto S(t)^{\tau-2} s_0(t)^{1+\beta-\tau}\ll s_0(t)^{\beta-1}\propto t^{-1} 
\end{equation}
since $\tau<2$ and $S(t)\gg s_0(t)$.  Therefore, $M_\beta$, being equivalent
to a power law $\ll 1/t$, is integrable. 

If $\tau=1+\beta$, from Eq. (\ref{logscal}),
\begin{equation}
M_\beta \propto S(t)^{\beta-1} \ln(S/s_0) \propto t^\frac{\beta-1}{1-\lambda} \ln t
\end{equation}
which is integrable when $\lambda>\beta$.
\vspace{0.2cm}

{\it (ii)} Now, let us consider the case when $M_1(t)$ {\it diverges at large time}.
From Eqs. (\ref{Mpoint}), (\ref{Mscalnorm}), (\ref{anormscal}) and (\ref{logscal}), we
obtain,
\begin{equation}\label{dotm1scal}
\dot{M_1}\propto \left\{ 
	\begin{array}{lr}
	   M_1 S^{\beta-1},& \qquad \mbox{if } \tau <1+\beta \\
	   M_1 S^{\tau-2} s_0(t)^{1+\beta-\tau}, &  \qquad \mbox{if } \tau
>1+\beta\\
	   M_1 S^{\beta-1}\ln(S/s_0),& \qquad \mbox{if } \tau=1+\beta
	\end{array}\right.
\end{equation} 
Anyway, in the three cases, $S\gg s_0$ implies that $\dot{M_1}\ll M_1
s_0^{\beta-1}$, hence, 
\begin{equation}\label{M1/M1}
\dot{M_1}\ll \frac{M_1}{t}
\end{equation}
Therefore $M_1(t)\ll t^\alpha$ for any $\alpha>0$.
From Eq. (\ref{Mscalnorm}) and the fact that $\dot{S}/{S}$ is at least of
order $1/t$ since $S(t)\gg t^{1/(1-\beta)}$, Eq. (\ref{M1/M1}) requires that,
\begin{equation}
2\frac{\dot{S}}{S}\sim \frac{\dot{Y}}{Y}
\end{equation} 
Thus, the scaling condition between (\ref{1scal}) and (\ref{3scal}) is
simply,
\begin{equation}
\frac{\dot{S}}{S} \propto \frac{S^{1+\lambda}}{Y} \propto M_1 S^{\lambda-1}
\end{equation} 
which implies that $S^{1-\lambda}$ is dominated by a power law.
Combined with the fact that $S(t)\gg t^{1/(1-\beta)}$, this requires that
\begin{equation}
\frac{\dot{S}}{S}\propto \frac{1}{t}
\end{equation} 
and 
\begin{equation}\label{M1S}
M_1\propto t^{-1}S^{1-\lambda}\gg t^{\frac{\beta-\lambda}{1-\beta}}
\end{equation}
thus, from Eq. (\ref{M1/M1}), we must have $\lambda\geq \beta$.
Now, combining    Eqs. (\ref{M1S}) and (\ref{dotm1scal}), we see that,
\begin{equation} \label{alpha123}
\dot{M_1}\propto M_1^{-\alpha_1} t^{-\alpha_2}
(\ln(t^\frac{\lambda-\beta}{1-\beta} M_1))^{\alpha_3}
\end{equation}
($\alpha_3=1$ if $\tau=1+\beta$, otherwise $\alpha_3=0$).

Since $M_1\to \infty$, the right hand side of the latter equation must 
be non integrable, and as $M_1$ is much smaller 
than any positive power of $t$, this
implies that $\alpha_2 \leq 1$.
\begin{equation}
\alpha_2= \left\{ 
	\begin{array}{ll}
	   \frac{1-\beta}{1-\lambda},& \qquad \mbox{if } \lambda \leq \beta \\
	   \frac{2-\tau}{1-\lambda}+ \frac{\tau-1-\beta}{1-\beta},&
  \qquad \mbox{if } \tau>1+\beta
	\end{array}\right. .
\end{equation}
Since $\tau\leq 1+\lambda$, $\alpha_2>1$ if $\lambda>\beta$. Therefore we
must have $\lambda\leq \beta$. However we already found that $\lambda\geq
\beta$, thus $\lambda$ must be equal to $\beta$.

As a consequence $\tau$ is never bigger than $1+\beta=1+\lambda$. 
and we can distinguish between $\mu>0$ kernels, for which
$\tau=1+\lambda=1+\beta$ and $\mu\leq 0$ kernels for which there is 
no polydispersity exponent or $\tau<1+\beta$.

Let us start with $\mu\leq 0$ kernels. As $\tau<1+\beta$, and
$\lambda=\beta$, Eq. (\ref{alpha123}) is reduced to  $\dot{M_1}\propto 1/t$,
which leads to,

\begin{equation}
M_1(t)\propto \ln t. 
\end{equation}

For a $\mu>0$ kernel, $\tau=1+\lambda=1+\beta$, and Eq. (\ref{alpha123}) 
leads to,
\begin{equation}
\dot{M_1} \propto (\ln M_1)/t,
\end{equation} 
and it is easily seen that, 
\begin{equation}
M_1(t)\propto (\ln t) \ln (\ln t).
\end{equation}

In both cases, we have,
\begin{eqnarray}
S(t)&\propto& (tM_1)^{\frac{1}{1-\beta}}, \\
Y(t)&\propto& S^2/M_1.  
\end{eqnarray}

Thus, the initial assumption that $S(t)\gg s_0(t)$ implies that 
$\lambda\geq \beta$ and that the scaling equation is Eq. (\ref{smolscal})
with $a=2b$ (since $2\dot{S}/{S}\sim \dot{Y}/{Y}$) and $0<a<+\infty$, and is
 the same as for standard Smoluchowski's equation with the same
kernel.

\vspace{0.5cm}

\subsubsection{Case $S\propto s_0$}
Conversely, let us assume that $S(t)\propto t^{1/(1-\beta)}$.
If $\dot{Y}/{Y}\gg \dot{S}/{S}\propto 1/t$, $Y$ increases faster than any
power law, the growth term is still negligible at large time, and Eqs.
(\ref{1scal}) and (\ref{3scal}) are of the same order, 
hence $S^{1+\lambda}/Y \propto \dot{Y}/{Y}$, which is contradictory, for 
$S^{1+\lambda}/Y$ vanishes faster than any power law while 
$\dot{Y}/{Y}\gg 1/t$.  

Thus, $\dot{Y}/{Y}=O(\dot{S}/{S})$ and both  terms in the 
right hand side of Smoluchowski's equation are of the
same order at large time, as $\dot{S}/{S}\propto S^{\beta-1}$.
 Besides, both terms must scale as the collision term,
otherwise the obtained scaling equation has no physical solution 
vanishing below a finite argument $x_0>0$. Thus,  (\ref{2scal}) and 
(\ref{3scal}) must be of the same order, which yields:
\begin{equation}
\frac{S^{\beta-1}}{Y}\propto \frac{S^{1+\lambda}}{Y^2}.
\end{equation}
and $Y(t)\propto S(t)^{2+\lambda-\beta}$.
The fact that the scaling function
vanishes at a finite $x_0>0$ ensures that, $M_1(t)\propto S^2/Y$,
hence,
\begin{equation}\label{m1stand}
 M_1(t)\propto S(t)^{\beta-\lambda}.
\end{equation}
Since $M_1(t)$ is non decreasing, we must have $\lambda\leq \beta$.
However, if $\lambda=\beta$, Eq. (\ref{massevol}) yields
\begin{equation}
\dot{M_1}\propto 1/t,
\end{equation}
hence $M_1(t)\propto \ln t$, which is in contradiction with
Eq. (\ref{m1stand}). Therefore one must have $\lambda<\beta$. 

The scaling equation has the form, 
\bleq
\ifpreprintsty\else
\renewcommand{\thesection}{\Alph{section}} %
\renewcommand{\thesubsubsection}{\alph{subsubsection}} %
\renewcommand{\thesubsection}{\arabic{subsection}} %
\renewcommand{\theequation}{\Alph{section}\arabic{equation}} \fi %
\begin{eqnarray} \label{bimod}
 b\left[\theta f(x) + xf'(x)\right]- a(x^\beta f(x))'&=& 
f(x)\int_0^{+\infty} f(x_1)K(x,x_1)dx_1\nonumber\\
& &- \frac{1}{2}\int_0^x f(x_1)f(x-x_1)K(x_1,x-x_1) \,dx_1
\end{eqnarray}
\eleq

\subsubsection{Conclusion}
Since $S(t)\geq s_0(t)$, the collection of the two cases we examined above
leads to the conclusion that, if $\lambda<1$ and $\beta<1$, there are three
main regimes of scaling, in agreement with the qualitative discussion in Sec
\ref{sec:scalth}.

If $\beta>\lambda$,  $S(t)$ scales as $S_g(t)\propto t^{1/(1-\beta)}$,
$Y(t)\propto S(t)^\theta$, with $\theta=2+\lambda-\beta$, and there is no
polydispersity exponent since the scaling function is zero below a finite
$x_0$.

If $\beta<\lambda$, $S(t)$ scales as $S_c(t)\propto t^{1/(1-\lambda)}$,
the mass is asymptotically conserved, i.e. $M_1(t)$ tends to a constant,
and $\theta=2$. There can be a polydispersity exponent, which is the same 
as for standard Smoluchowski's equation Eq. (\ref{smoleq}) 
with the same kernel.

Eventually, in the marginal case when $\lambda=\beta$, the scaling of $S(t)$
 depends on the kernel not only through its homogeneity $\lambda$, but also
through its $\mu$ exponent. As in the $\beta<\lambda$ case, the scaling
equation is the same as for Smoluchowski's equation with the same kernel.
For $\mu\leq 0$ kernels, the mass in the system $M_1(t) \propto \ln t$,
while for $\mu>0$ kernels, with $\tau=1+\lambda$, $M_1(t)\propto t (\ln t)
\ln(\ln t)$. In both cases, $S\propto (tM_1)^{1/(1-\beta)}$ and $Y\propto S^2/M_1$.

 These scaling results can be compared to  the case $K=1$,$\beta=0$ which 
we solved exactly: we found that the conventional scaling breaks down and
that, with the proper scaling form, the scaling function   is the same
as for the exactly solvable standard Smoluchowski's equation without the growth
term, that $S(t) \propto (t\ln t)$, $M_1(t)\propto\ln t$ and $Y\propto 
t^2\ln t$, just as predicted by the scaling theory.

\subsection{$\lambda=1$ and $\beta<1$}
For $\lambda=1$ and $\beta <1$, it is possible to follow the same line
of reasoning, with a few modifications.
In this case, one has to distinguish between $\mu>0$ and $\mu\geq 0$ (this
is also true for standard Smoluchowski's equation with $\lambda=1$ 
\cite{vdg87scal}),
since for $\mu>0$, we find $\tau=2$ and the scaling of $M_1$ has an extra
$\ln (S/s_0)$. It is found that $M_1$ is asymptotically conserved, that 
$S(t)\gg s_0(t)$, and that the scaling equation is Eq. (\ref{smolscal}) with
$a=2b<+\infty$.   
For $\mu>0$, one has $\dot{S} \propto S/(\ln S)$, which
leads to, 
\begin{equation}
	 S(t)\propto e^{b\sqrt{t}},
\end{equation}
whereas if $\mu\leq 0$,
\begin{equation}
S(t)\propto e^{bt},
\end{equation}
where $b$ cannot be derived from the scaling theory.

\subsection{$\beta=1$ and $\lambda<1$}
In this case $s_0(t)\propto e^t$, and the discussion is quite different.
From Eq. (\ref{Mpoint}), we see that:
\begin{equation}
M_1(t)=M_1(0)\,e^t.
\end{equation}
Since $S(t)\geq s_0(t)\propto e^t$, we have in the large time 
limit $\dot{S}/S \geq 1$. Let us assume that $\dot{S}/{S}\gg 1$, i.e.
that $S(t)$ is bigger than any exponential function $e^{\alpha t}$,
which entails that
 (\ref{1scal}) is much bigger than (\ref{2scal}) (which scales as $1/Y$
since $\beta=1$).
 From Eqs. (\ref{scalnorm}), (\ref{anormscal}) and (\ref{logscal}),
it is clear that, 
\begin{equation}\label{mmajore}
S^2/Y= O(M_1(t))= O(e^t).
\end{equation} 
Consequently, if   (\ref{1scal}) scales as (\ref{3scal}), then 
$\dot{S}/{S^\lambda}=O(e^t)$ and $\dot{Y}/{Y^{(1+\lambda)/2}}=O(e^{t/2})$.
Since $\lambda<1$, these two relations are in contradiction with the 
assumption that $S$ is much bigger than any exponential function (which 
implies the same property for $Y$, through Eq. (\ref{mmajore})).
 We see that if $\dot{S}/{S}\gg 1$ , (\ref{1scal}) is the leading term 
in the scaling limit, and the scaling function is a pure power law
$f(x)=c x^{-\tau}$ with $\tau= \lim (\dot{Y}/Y)(S/\dot{S})$. 	One must
have $\tau>2$ such that the total mass in the system be finite at finite
time in the scaling regime. Making use of Eq. (\ref{anormscal}),
we find that $n(t)\propto M_1(t)/s_0(t)$ would tend to a finite value
$n_\infty>0$, which is unphysical.

Therefore $\dot{S}/{S}$ is of order $1$. From arguments very similar 
to those we just used, it is easily seen that 
$\dot{Y}/{Y}$ cannot be much bigger than $1$. Thus, (\ref{1scal}) scales as 
(\ref{2scal}). If $\dot{Y}/Y \to 1$ and $\dot{S}/S
\to 1$, the left hand side of Eq. (\ref{gensmol}) vanishes and we have to take into
account the subleading terms in the scaling limit. This occurs for
$\lambda=0$ as  will be seen below. 

If the left hand side does not vanish, 
  the scaling with (\ref{3scal}) leads to $S^{1+\lambda}\propto
Y$,  and the scaling equation is once again Eq. (\ref{smolscal}), but now with
$b/a= (1-\dot{S}/{S})/(1-\dot{Y}/{Y})$.  Consequently, the polydispersity
exponent, if any, is less than $1+\lambda$, and Eq. (\ref{Mscalnorm}) holds,
leading to  $S^2\propto Y\, e^t$. Since $S^{1+\lambda}\propto Y$, we have,
\begin{eqnarray}
S(t)&\propto& e^\frac{t}{1-\lambda}\\
Y(t)&\propto& S^2 \, e^{-t}
\end{eqnarray}
which excludes $\lambda <0$ since $S(t)\geq s_0(t)=e^t$, and also
$\lambda=0$ for which $\dot{S}/S\to 1$ and $\dot{Y}/Y \to 1$.
Note that in the $\lambda>0$ case, we find $a=2b$, and once again the scaling
equation is the same as for standard Smoluchowski's equation.

Indeed, for the exactly solvable case $K=1,\, \beta=1$, which corresponds
to $\lambda=0$, we found that $S(t)\propto te^t$ and $Y(t)\propto S^2/e^t$,
thus $\dot{S}/S\to 1 $ and $\dot{Y}/Y\to 1$.  
Thus, to treat the $\lambda=0$ case, we  shall write $S(t)=X(t)M_1(t)$,
 with $\dot{X}/X \ll 1$, and we have $Y\propto S^2/M_1=XS$. 
The right hand side of Eq. (\ref{gensmol}) scales as,
\begin{equation}
 -\frac{1}{Y} (2f(x)+xf'(x)) \frac{\dot{X}}{X},
\end{equation} 
while the left hand side scales
as 
\begin{equation}
\frac{S}{Y^2}(..)\propto \frac{X}{Y} (..) 
\end{equation} which leads to $X(t)\propto t$, recovering the
exact result for $K=1$. Once again the scaling function is  Eq. 
(\ref{smolscal}) with $a=2b<+\infty$. The polydispersity exponent $\tau$ is
strictly less than $2$, which justifies a posteriori that $Y\propto
S^2/M_1$ (it is possible to show that assuming $\tau>2$ leads to a 
contradiction).

However, for $\lambda<0$, we were unable to find a consistent scaling.

\subsection{$\lambda=1$ and $\beta=1$}
In this case, we still have $M_1(t)\propto e^t$ and $s_0(t)\sim e^t$, but
it is easily seen with the same kind of arguments as above, that one must 
have $\dot{S}/{S}\gg 1$. Thus the exogenous growth term (\ref{2scal}) is 
negligible, and the scaling of (\ref{1scal}) and (\ref{2scal}) yields,
$\dot{S}/S\propto S^2/Y$.

For $\mu\leq 0$ kernels, one has $M_1\propto S^2/Y$ and we obtain 
$\dot{S}/S\propto e^t$, leading to,
\begin{equation}
S(t)\propto e^{be^t}.
\end{equation}
For $\mu>0$, we have $\tau=2$ and $M_1\propto S^2\ln (S/e^t)/Y$, leading to
$\dot{S}/S\propto e^t/\ln(S)$, and,
\begin{equation}
S(t)\propto e^{b\sqrt{e^t}}.
\end{equation}
In these expressions $b$ is an unknown positive constant.
\bibliography{}

\begin{thebibliography}{10}

\bibitem{friedlander77}
S.~K. Friedlander,
\newblock {\em Smoke, dust and haze},
\newblock Wiley Interscience, New York, 1977.

\bibitem{family84}
F.~Family and D.~Landau, editors,
\newblock {\em Kinetics of Aggregation and Gelation},
\newblock North Holland,Amsterdam, 1984.

\bibitem{stanley86}
H.~E. Stanley and N.~Ostrowsky, editors,
\newblock {\em On Growth and Form},
\newblock Martinus Nijhoff, 1986.

\bibitem{vicsek92}
T.~Vicsek,
\newblock {\em Fractal growth phenomena},
\newblock World Scientific, Singapore, second edition, 1992.

\bibitem{meakindrop92}
P.~Meakin,
\newblock Rep. Prog. Phys. {\bf 55}, 157 (1992).

\bibitem{beysens86}
D.~Beysens and C.~Knobler,
\newblock Phys. Rev. Lett. {\bf 57}, 1433 (1986).

\bibitem{family88}
F.~Family and P.~Meakin,
\newblock Phys. Rev. Lett. {\bf 61}, 428 (1988).

\bibitem{family89}
F.~Family and P.~Meakin,
\newblock Phys. Rev. A {\bf 40}, 3836 (1989).

\bibitem{fritter91}
D.~Fritter, C.~Knobler, and D.~Beysens,
\newblock Phys. Rev. A {\bf 43}, 2858 (1991).

\bibitem{derrida91}
B.~Derrida, C.~Godr\`eche, and I.~Yekutieli,
\newblock Phys. Rev. A {\bf 44}, 6241 (1991).

\bibitem{marcos95}
M.~Marcos-Martin, D.~Beysens, J.-P. Bouchaud, C.~Godr\`eche, and I.~Yekutieli,
\newblock Physica A {\bf 214}, 396 (1995).

\bibitem{klett75}
J.~Klett,
\newblock J. Atmos. Sci {\bf 32}, 380 (1975).

\bibitem{white82}
W.~H. White,
\newblock J. Colloid Interface Sci. {\bf 87}, 204 (1982).

\bibitem{crump82}
J.~G. Crump and J.~H. Seinfeld,
\newblock J. Colloid Interface Sci. {\bf 90}, 469 (1982).

\bibitem{hendriks85}
E.~M. Hendriks and R.~M. Ziff,
\newblock J. Colloid Interface Sci. {\bf 105}, 247 (1985).

\bibitem{vicsek85}
T.~Vicsek, P.~Meakin, and F.~Family,
\newblock Phys. Rev. A {\bf 32}, 1122 (1985).

\bibitem{racz85}
Z.~R\'acz,
\newblock Phys. Rev. A {\bf 32}, 1129 (1985).

\bibitem{hayakawa87}
H.~Hayakawa,
\newblock J. Phys. A {\bf 20}, L801 (1987).

\bibitem{takayasu88}
H.~Takayasu, I.~Nishikawa, and H.~Tasaki,
\newblock Phys. Rev. A {\bf 37}, 3110 (1988).

\bibitem{takayasu89}
H.~Takayasu,
\newblock Phys. Rev. Lett. {\bf 63}, 2563 (1989).

\bibitem{majumdar93}
S.~N. Majumdar and C.~Sire,
\newblock Phys. Rev. Lett. {\bf 71}, 3729 (1993).

\bibitem{smoluchowski18}
M.~von Smoluchowski,
\newblock Z. Phys. Chem {\bf 92}, 129 (1918).

\bibitem{vdg89}
P.~G.~J. van Dongen,
\newblock Phys. Rev. Lett. {\bf 63}, 1281 (1989).

\bibitem{vdg85prl}
P.~G.~J. van Dongen and M.~H. Ernst,
\newblock Phys. Rev. Lett. {\bf 54}, 1396 (1985).

\bibitem{vdg87scal}
P.~G.~J. van Dongen and M.~H. Ernst,
\newblock J. Stat. Phys. {\bf 50}, 295 (1987).

\bibitem{nontriv97}
S.~Cueille and C.~Sire,
\newblock Phys. Rev. E {\bf 55}, 5465 (1997).

\bibitem{droplett}
S.~Cueille and C.~Sire,
\newblock Smoluchowski's equation for cluster exogenous growth,
\newblock submitted to Europhys. Lett., 1997.

\bibitem{vincent71}
R.~Vincent,
\newblock Proc. Roy. Soc. A {\bf 321}, 53 (1971).

\bibitem{krapivsky96}
P.~Krapivsky and S.~Redner,
\newblock Phys. Rev. E {\bf 54}, 3553 (1996).

\bibitem{ziff84}
R.~M. Ziff, M.~H. Ernst, and E.~M. Hendriks,
\newblock J. Colloid Interface Sci. {\bf 100}, 220 (1984).

\bibitem{ernst84}
M.~H. Ernst, R.~M. Ziff, and E.~M. Hendriks,
\newblock J. Colloid Interface Sci. {\bf 97}, 266 (1984).

\bibitem{leyvraz82}
F.~Leyvraz and H.~Tschudi,
\newblock J. Phys. A {\bf 15}, 1951 (1982).

\bibitem{hendriks83}
E.~Hendriks, M.~Ernst, and R.~Ziff,
\newblock J. Stat. Phys. {\bf 31}, 519 (1983).

\bibitem{kang86}
K.~Kang, S.~Redner, P.~Meakin, and F.~Leyvraz,
\newblock Phys. Rev. A {\bf 33}, 1171 (1986).

\bibitem{vdg85jpa}
P.~G.~J. van Dongen and M.~H. Ernst,
\newblock J. Phys. A {\bf 18}, 2779 (1985).

\bibitem{krivitsky95}
D.~S. Krivitsky,
\newblock J. Phys. A {\bf 28}, 2025 (1995).

\bibitem{song}
S.~Song and D.~Poland,
\newblock Phys. Rev. A {\bf 46}, 5063 (1992).

\bibitem{meakin83}
P.~Meakin,
\newblock Phys. Rev. Lett. {\bf 51}, 1119 (1983).

\bibitem{kolb83}
M.~Kolb, R.~Botet, and R.~Jullien,
\newblock Phys. Rev. Lett. {\bf 51}, 1123 (1983).

\bibitem{meakinrev92}
P.~Meakin,
\newblock Physica Scripta {\bf 46}, 46 (1992).

\bibitem{mcmurry80}
P.~H. McMurry,
\newblock J. Colloid Interface Sci. {\bf 78}, 513 (1980).

\bibitem{hendriks84}
E.~M. Hendriks,
\newblock J. Phys. A {\bf 17}, 2299 (1984).

\bibitem{bak87}
P.~Bak and K.~Wiesenfeld,
\newblock Phys. Rev. Lett. {\bf 59}, 381 (1987).

\bibitem{dhar90}
D.~Dhar,
\newblock Phys. Rev. Lett. {\bf 64}, 1613 (1990).

\bibitem{tanaka75}
H.~Tanaka,
\newblock J. Heat Transfer {\bf 97}, 72 (1975).

\bibitem{meakindyn90}
P.~Meakin,
\newblock Simple models for coalescence of fluid droplets,
\newblock in {\em Dynamics and patterns in complex fluids}, edited by A.~Onuki
  and K.~Kawasaki, Springer Proceedings in Physics 52, 1990.

\end{thebibliography}
\ecols

\end{document}